\documentstyle[12pt,epsf]{article}
\epsfverbosetrue
\def\figlabel#1{\xdef#1{\thefigure}}

\def\figalign#1#2#3#4#5#6{
\begin{figure}
\centerline{
\hbox to 2.5truein{\vtop{\hsize=2.5truein\epsfxsize=6cm
\centerline{\epsfbox{#1} }
\caption[]{#3}
\figlabel{#2} }}
\qquad\hbox to 2.5truein{\vtop{\hsize=2.5truein\epsfxsize=6cm
\centerline{\epsfbox{#4} }
\caption[]{#6}
\figlabel{#5} }} }
\end{figure} }

\def\be{\begin{equation}}
\def\ee{\end{equation}}
\def\bea{\begin{eqnarray}}
\def\eea{\end{eqnarray}}

\begin{document}
\begin{titlepage}
\begin{flushright} { ~}\vskip -1in CERN-TH/97-360\\ US-FT-34/97\\
hep-th/9712139\\ 
\end{flushright}
\vspace*{20pt}
\bigskip
\begin{center}
 {\Large Gauge-Invariant Operators for Singular Knots in Chern-Simons
Gauge Theory}
\vskip 0.9truecm

{J. M. F. Labastida$^{a,b}$ and  Esther P\'erez$^{b}$}

\vspace{1pc}

{\em $^a$ Theory Division, CERN,\\
 CH-1211 Geneva 23, Switzerland.\\
 \bigskip
  $^b$ Departamento de F\'\i sica de Part\'\i culas,\\ Universidade de
Santiago de Compostela,\\ E-15706 Santiago de Compostela, Spain.\\}

\vspace{5pc}

{\large \bf Abstract}
\end{center} We construct gauge invariant operators for singular knots in
the context of Chern-Simons gauge theory. These new operators provide
polynomial invariants and Vassiliev invariants for singular knots. As an
application we present the form of the Kontsevich integral for the case of
singular knots.

\vspace{6pc}

\begin{flushleft} { ~}\vskip -1in CERN-TH/97-360\\ December 1997\\
\end{flushleft}


\end{titlepage}

\def\theequation{\thesection.\arabic{equation}}

\def\del{{\delta^{\hbox{\sevenrm B}}}} \def\ex{{\hbox{\rm e}}}
\def\azb{A_{\bar z}} \def\az{A_z} \def\bzb{B_{\bar z}} \def\bz{B_z}
\def\czb{C_{\bar z}} \def\cz{C_z} \def\dzb{D_{\bar z}} \def\dz{D_z}
\def\im{{\hbox{\rm Im}}} \def\mod{{\hbox{\rm mod}}} \def\tr{{\hbox{\rm
Tr}}}
\def\ch{{\hbox{\rm ch}}} \def\imp{{\hbox{\sevenrm Im}}}
\def\trp{{\hbox{\sevenrm Tr}}} \def\vol{{\hbox{\rm Vol}}}
\def\rl{\Lambda_{\hbox{\sevenrm R}}} \def\wl{\Lambda_{\hbox{\sevenrm W}}}
\def\fc{{\cal F}_{k+\cox}} \def\vev{vacuum expectation value}
\def\nodiv{\mid{\hbox{\hskip-7.8pt/}}}
\def\ie{{\em i.e.}}

\def\np{Nucl. Phys.}
\def\pl{Phys. Lett.}
\def\prl{Phys. Rev. Lett.}
\def\pr{Phys. Rev.}
\def\ap{Ann. Phys.}
\def\cmp{Comm. Math. Phys.}
\def\ijmp{Int. J. Mod. Phys.}
\def\jmp{J. Math. Phys.}
\def\mpl{Mod. Phys. Lett.}
\def\inma{Invent. Math.}
\def\tam{Trans. Am. Math. Soc.}
\def\lmp{Lett. Math. Phys.}
\def\bams{Bull. AMS}
\def\am{Ann. of Math.}
\def\rmp{Rev. Mod. Phys.}
\def\jpsc{J. Phys. Soc. Jap.}
\def\topo{Topology}
\def\kjm{Kobe J. Math.}
\def\knot{Journal of Knot Theory and Its Ramifications}

\newcommand{\ZZ}{{\mbox{{\bf Z}}}}
\newcommand{\RR}{{\mbox{{\bf R}}}}
\newcommand{\MM}{{\mbox{{\bf M}}}}
\newcommand{\CC}{{\mbox{{\bf C}}}}
\newcommand{\RRs}{{\scriptstyle {\rm {\bf R}}}}
\newcommand{\MMs}{{\scriptstyle {\rm {\bf M}}}}
\newcommand{\CS}{{\scriptstyle {\rm CS}}}
\newcommand{\CSs}{{\scriptscriptstyle {\rm CS}}}
\newcommand{\beq}{\begin{equation}}
\newcommand{\eeq}{\end{equation}}
\newcommand{\bear}{\begin{eqnarray}}
\newcommand{\eear}{\end{eqnarray}}
\newcommand{\W}{{\cal W}}
\newcommand{\F}{{\cal F}}
\newcommand{\x}{{\cal O}}\newcommand{\LL}{{\cal L}}

\def\mani{{\cal M}}
\def\calo{{\cal O}}
\def\calb{{\cal B}}
\def\calw{{\cal W}}
\def\calz{{\cal Z}}
\def\cald{{\cal D}}
\def\calc{{\cal C}}
\def\to{\rightarrow}
\def\ele{{\hbox{\sevenrm L}}}
\def\ere{{\hbox{\sevenrm R}}}
\def\zb{{\bar z}}
\def\wb{{\bar w}}
\def\nodiv{\mid{\hbox{\hskip-7.8pt/}}}
\def\menos{\hbox{\hskip-2.9pt}}
\def\dr{\dot R_}
\def\drr{\dot r_}
\def\ds{\dot s_}
\def\da{\dot A_}\def\dga{\dot \gamma_}
\def\ga{\gamma_}
\def\dal{\dot\alpha_}
\def\al{\alpha_}
\def\cl{{\it closed}}
\def\cls{{\it closing}}
\def\vev{vacuum expectation value}
\def\tr{{\rm Tr}}
\def\to{\rightarrow}
\def\too{\longrightarrow}


\newfont{\namefont}{cmr10}
\newfont{\addfont}{cmti7 scaled 1440}
\newfont{\headfontb}{cmbx10 scaled 1728}
%

\section{Introduction}
\setcounter{equation}{0}

Chern-Simons gauge theory \cite{csgt} has shown to be a very powerful tool
to study knot and links invariants. Its analysis from both, the
perturbative and the non-perturbative points of view has provided
numerous important insights in the theory of these invariants.
Non-perturbative methods \cite{csgt,nbos,torus,king,martin,kaul} have
established the connection of Chern-Simons gauge theory with polynomial
invariants as the Jones polynomial \cite{jones} and its generalizations
\cite{homfly,kauffman,aku}. Perturbative methods
\cite{gmm,natan,vande,alla,torusknots,alts,lcone} have provided
representations of Vassiliev invariants.

In the context of Chern-Simons gauge theory, knot and link invariants have
been obtained studying the vacuum expectation values of the associated
Wilson loops, or product of Wilson loops. Since these loops correspond to
knots or components of links, there are no intersections and the vacuum
expectation values are well defined. Indeed, these vacuum expectation
values turn out to be knot and link invariants. The story is different if
one considers vacuum expectation values of Wilson loops or products of
Wilson loops with intersections. The presence of singularities complicates
the analysis and it is not known if they can be defined consistently.
Fortunately, these objects are not the ones in which one is interested in
the study of singular knots and links from the point of view of the theory
of Vassiliev invariants. In this theory one must regard the invariant
associated to a knot or link with an intersection as a certain difference
of the invariants which correspond to different ways of solving the
intersection. For example, if one considers the case of knots with a
double point, invariants associated to the singular knot are defined as a
difference between the invariant associated to the two resolutions of the
double point (overcrossing and undercrossing). Invariants for singular
knots with more than one double point are defined following an iterative
procedure. The problem which we address in this paper is the construction
of the gauge invariant operators associated to singular knots when their
invariants are defined as differences of invariants involving singular
knots with one less double point.

Singular knots have been studied in the context of Chern-Simons gauge
theory to build physical states of quantum gravity in the Ashtekar
formalism \cite{gambinidos}. Our work can be regarded in part as a
development of the approach to invariants for singular knots initiated in
\cite{gambinidos}. Singular knots have been also considered in other
works \cite{brug,gambini}, as they appear as intermediate knot
configurations in the study of non-singular knot invariants from a
perturbative point of view. In the present work the emphasis is focused
on the study of the singular knots themselves and on the construction of
a generalization of the Wilson loop which leads to singular knots
invariants satisfying the features imposed by the theory of Vassiliev
invariants.

Our construction is based on functional integral technics. It involves the
use of standard formulae for the variation a Wilson line under a
deformation of its path. We will analyze  families of singular knots
parametrized by a parameter $u$ such that for $u=0$ there occurs a
self-intersection, a double point,  at some point $P\in M$, where $M$ is some boundaryless smooth three-manifold. For
$u\neq 0$ there are not self-intersections in a neighborhood of $P$. If
the singular knots of a given family are parametrized by a parameter
$v$,  two values of this parameter, $v=s$ and $v=t$, with
$s\neq t$, correspond to the point $P$ for $u=0$. All singular knots with
$u>0$ are equivalent under rigid vertex  isotopy. The same property holds
for all the ones with $u<0$. The difference between the two sets is that
an undercrossing near $P$ is replaced by an overcrossing. A review on the basic concepts involved in the theory of singular knots can be found in
\cite{bilin}. If we denote by
$\psi_+$ the invariant associated to the singular knot for  $u>0$, and by
$\psi_-$ the one for $u<0$, the invariant associated to the singular knot
with one more double point is defined as $\psi_+ - \psi_-$. In our
construction we will analyze this difference computing the first
derivative respect to
$u$ of the vacuum expectation value of the operator associated to the
family of singular knots. As a function of $u$ such a vacuum expectation
value  behaves as a step function around $u=0$. We will indeed find that
its first derivative respect to $u$ has a delta-function behavior. Its
integration along a finite interval enclosing $u=0$ is then carried out
very simply, providing the searched operators whose vacuum expectation
values lead to
$\psi_+ - \psi_-$.

\begin{figure}
\centerline{\hskip.4in \epsffile{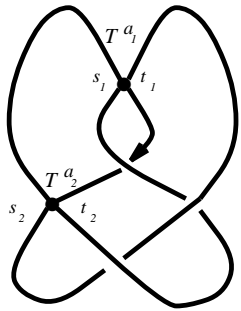}}
\caption{Example of a singular knot with two double points.}
\label{singu}
\end{figure}

The form of the new operators turn out to be very simple. Let us consider
a singular oriented knot $K^n$ (parametrized by a parameter $v$) with
$n$ singularities, all of them consisting of double points. These occur at
points $P_i$,
$i=1,\dots,n$, in $M$. Let us choose as base point one of them and let us
denote by $s_i$, $t_i$, with $s_i<t_i$, for $i=1,\dots,n$, the values of
the parameter $v$  at the double points. In addition, to each double point
we also associate a group generator whose group index is labeled as
$T^{a_i}$,
$i=1,\dots,n$. Thus, at each double point $P_i$ we have a triple
$\tau_i=\{s_i,t_i,T^{a_i}\}$. Traveling along the knot starting from the base
point the parameter $v$  takes the value $s_i$  when one encounters the
singular point
$P_i$ for first time, and $t_i$ when one encounters it a second time. All
the elements of the set $\{s_i,t_i; i=1,\dots,n\}$  are different. We can
therefore order them into $\{w_1,w_2,\dots,w_n\}$ with $w_j < w_{j+1}$.
Each $w_j$ belongs to one and only one of the triples $\tau_i$. This order
induces a mapping $\phi(w_j)$ which assigns to $w_j$ the index of the
group generator in the triple to which $w_j$ belongs. The gauge invariant
operator associated to the singular knot $K^n$ has the form:
\bear && {\hskip-2cm} (2x)^n \tr \Big[T^{\phi(w_1)} U(w_1,w_2)
T^{\phi(w_2)} U(w_2,w_3) T^{\phi(w_3)}\cdots \nonumber\\ && 
{\hskip3cm}
\cdots
U(w_{2n-1},w_{2n}) T^{\phi(w_{2n})} U(w_{2n},w_{1}) \Big],
\nonumber\\
\label{operador}
\eear where,
\beq x = {2\pi i \over k},
\label{laequis}
\eeq and $U(w_j,w_{j+1})$ denotes the Wilson line along the path joining
the points corresponding to $v=w_j$ and $v=w_{j+1}$. Notice that the
mapping $\phi$ is two to one and therefore there are not free group
indices. Notice also that (\ref{operador}) does not depend on the choice
of base point since it is invariant under cyclic permutations of the
singular points.

\begin{figure}
\centerline{\hskip.4in \epsffile{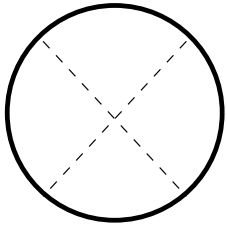}}
\caption{Configuration corresponding to the singular knot shown in fig. 1.}
\label{confi}
\end{figure}

The structure of the operators (\ref{operador}) allows to classify the
singular knots with $n$ double points into classes. Given a singular knot
with $n$ double points we define its configuration as the diagram which
results after drawing a circle and marking $2n$ points on it in the
order that the group generators appear in (\ref{operador}). The points
corresponding to generators with the same index are joined by dashed
line. The configuration corresponding to the singular knot shown in fig. \ref{singu} has been depicted in fig. \ref{confi}. Two singular knots are
in the same class if they possess the same configuration. This assignment
of configurations to singular knots, suggested by the operators
(\ref{operador}), is the same as the standard one used in knot theory
(see, for example, \cite{bilin}).

Though the general form of the operators (\ref{operador}) might look
somehow cumbersome, it is straightforward to write down the operator
(\ref{operador}) for specific cases. Let us illustrate this with an
example. Let us consider the singular knot depicted in fig. \ref{singu}.
One first assigns to each singular point the triple made by the two values
that the parameter $v$ takes and a group generator with some group index. Then one
chooses one of the double points as  base point and travels along the
knot  building a product of Wilson lines and generators as they are
encountered in the journey. For the oriented singular knot in fig.
\ref{singu} the resulting operator is:
\beq (2x)^2\tr  \Big[T^{a_1} U(s_1,s_2) T^{a_2} U(s_2,t_1) T^{a_1}U(t_1,t_2)
T^{a_2} U(t_2,s_1)  \Big].
\label{ejemplo}
\eeq

As we will show in the following section the operators (\ref{operador})
are gauge invariant. Their  form constitutes a field-theory-based simple
proof of the theorem by Birman and Lin \cite{bilin} which states that the
coefficient of order
$n$ of the expansion of a polynomial invariant (after replacing its
variable by $e^x$) is a Vassiliev invariant of order $n$. Indeed,
(\ref{operador}) shows that the perturbative expansion of a polynomial
invariant of an oriented knot with $n$ double points starts at order
$n$. All the coefficients of lower order have therefore vanished and
thus all these coefficients are Vassiliev invariants of order lower
than $n$. 

The expression (\ref{operador}) constitutes also a field-theory-based
simple proof  of the theorem by Bar-Natan \cite{barnatan} which states
that semi-simple Lie algebras are Vassiliev invariants for singular knots
which can be integrated to Vassiliev invariants for non-singular knots.
The zero-order of the perturbative expansion of the vacuum expectation
value of the operators (\ref{operador}), which is just the result of
setting
$U=1$, shows that semi-simple Lie algebras lead to Vassiliev invariants of
order
$n$ for oriented singular knots with $n$ double points.

These two examples constitute a sign of the wide variety of  applications
that the operators (\ref{operador}) have. In this paper we will study the
perturbative series expansion associated to the vacuum expectation value
of (\ref{operador}) in the light-cone gauge. As a result we obtain a
generalization of the Kontsevich integral for Vassiliev invariants
\cite{kont,tung} which is valid in the case of singular knots. We
will present also the general form of the Vassiliev invariant of order
$n$ for a singular knot with $n-1$ double points. This last result allows
to complete very simply the next to the top row of an actuality table.

The paper is organized as follows. In sect. 2 we present the construction
of the gauge invariant operators associated to singular knots. In sect. 3
we discuss some immediate consequences which follow very simply from the
form of these operators. In sect. 4 we analyze the new operators in the
light-cone gauge and we present a generalization of the Kontsevich
integral for ordinary knots to the case of singular knots. In sect. 5 we
carry out one more application: we present the general form of the
Vassiliev invariants of order $n$ for singular knots with $n-1$
singularities.  Finally, in sect. 6 we state our conclusions.

\vfill
\eject

\section{Operators for singular knots}
\setcounter{equation}{0}

In this section we construct the gauge invariant operators associated to
singular knots in the context of Chern-Simons gauge theory. We will first
review some standard properties of Wilson lines. Then, after the analysis
of the variation of their vacuum expectation value, we will carry out the
construction of the operators (\ref{operador}). 

Let us consider Chern-Simons gauge theory on a boundaryless smooth
three-manifold
$M$ for a semi-simple gauge group $G$. The action of the theory, $S(A)$,
is built out of the Chern-Simons three-form,
\beq  S (A)={k\over 4\pi}\int_{M} \tr  \Big(A\wedge dA +  {2\over 3}
A\wedge A\wedge A\Big),
\label{action}
\eeq where $A_\mu=A_\mu^a T^a$ is a $G$-connection, $k$ an integer
parameter, and Tr denotes the trace in the fundamental representation of
the gauge group
$G$. The quantities $T^a$, $a=1,\dots,$dim$G$, are the group generators,
which are normalized so that $\tr(T^aT^b)={1\over 2}\delta^{ab}$. One
salient property of the action (\ref{action}) is that its variation
respect to the gauge connection has the form:
\beq {\delta \over \delta A_\mu^a(x)} S (A) = {k\over 8\pi}
\epsilon^{\mu\nu\rho} F_{\nu\rho}^a(x),
\label{fieldequation}
\eeq where $F_{\nu\rho}$ is the curvature of $A_\mu$.

\begin{figure}
\centerline{\hskip.4in \epsffile{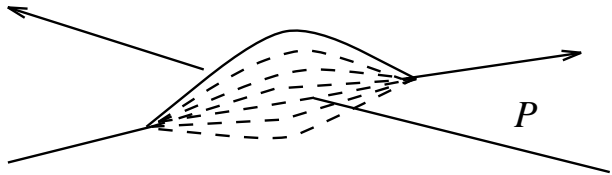}}
\caption{Resolution of a singular knot by a continuous of paths.}
\label{param}
\end{figure}

The natural observables of Chern-Simons gauge theory are Wilson loops and
graphs (or spin networks). Their vacuum expectation values are related to
knot, link and graph invariants. In this work we are interested in
analyzing the variation of the vacuum expectation values of Wilson loops
under a deformation of their associated path. Actually, we need to study
first a slightly more involved object: the Wilson line.  Let us first
recall its definition and review some of its properties. Let $\gamma$ be a
path in
$M$, \ie\ a smooth map $\gamma:[s,t]\rightarrow M$. The associated Wilson
line $U_\gamma(s,t)$ is defined as:
\beq U_\gamma(s,t) = {\rm P} \exp \int_s^t dv \,\dot\gamma(v) A(\gamma(v)),
\label{linea}
\eeq where P denotes path ordering. In (\ref{linea}) there is an implicit
choice of a representation of the gauge group. The vacuum expectation
value of a Wilson line is defined as the functional integral:
\beq
\Big\langle U_\gamma(s,t) \Big\rangle =
\int [DA] U_\gamma(s,t) \ex^{i S(A)}.
\label{valor}
\eeq

\begin{figure}
\centerline{\hskip.4in \epsffile{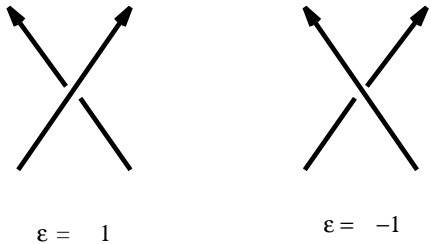}}
\caption{Signatures at the crossings.}
\label{signat}
\end{figure}

We will consider a family of smooth paths $\gamma_u$ parametrized by the
continuous parameter $u$, such that for $u=0$ the path$\gamma_0$ possesses
a self-intersection at some point $P\in M$, \ie\ for $u=0$ it has a 
double point. For
$u>0$ ($u<0$) the path presents an overcrossing (undercrossing) near the
point
$P$. A family of paths with these features has been pictured in fig.
\ref{param}. The path $\gamma_0(v)$ has a double point at $v=s_1$ and
$v=t_1$ with
$s_1<t_1$. Paths $\gamma_u(v)$ with $u\neq 0$ are different than
$\gamma_0(v)$ only in a region in parameter space around $v=s_1$. The
derivative of $\gamma_u(v)$ respect to $u$ is only non-zero in that
region. It vanishes away from $v=s_1$, in particular at $v=t_1$. In the
resolution of a double point we will call overcrossing (undercrossing) to
the one which leads to a crossing with positive (negative) sign, as
depicted in fig. \ref{signat}. Our goal is to study the first derivative
of the vacuum expectation value of the Wilson line
$U_{\gamma_u}(s,t)$ respect  to the parameter $u$. As stated in the
introduction, due to the topological character of Chern-Simons gauge
theory, one expects a step-function behavior for
$\langle U_{\gamma_u}(s,t) \rangle$ as a function of $u$ in a neighborhood
of
$u=0$. This implies the presence of a delta function in its derivative. As
we show below, this is in fact what we find.

Our starting point is the well known formula for the variation of a Wilson
line under a deformation of its path:
\beq {d\over d u} U_{\gamma_u}(s,t) =
\int_s^t dv\,  \gamma_u^{'\mu}(v) \dot\gamma_u^\nu(v)
   U_{\gamma_u}(s,v) F_{\mu\nu}(\gamma(v)) U_{\gamma_u}(v,t),
\label{variacion}
\eeq where we have denoted derivatives respect to the path parameter by a
dot, and derivatives respect to $u$ by a prime. Recall that
$\gamma_u^{'}(v)$ is only different from zero in a region in parameter
space around $v=s_1$.  Another important property of the Wilson line
which will be used later is its behavior under a functional derivation
respect to the gauge connection:
\beq {\delta \over \delta A^a_\mu(x)} U_{\gamma_u}(s,t) =
\int_s^t dw\, \dot\gamma_u^\mu(w) \delta^{(3)}(x,\gamma_u(w))  
U_{\gamma_u}(s,w) T^a U_{\gamma_u}(w,t).
\label{derivada}
\eeq

Taking into account (\ref{variacion}) and (\ref{fieldequation}) and
integrating by parts in connection space one can write the derivative
respect to $u$ of the vacuum expectation value (\ref{valor}) as:
\bear && {d\over du} \langle U_{\gamma_u}(s,t) \rangle= \nonumber\\ &&
{\hskip-0.8cm} {4\pi i \over k}\int [DA]
\ex^{i S(A)}
\int_s^t dv\, \epsilon_{\mu\nu\rho} \gamma_u^{'\mu}(v)\dot\gamma_u^\nu(v)
 {\delta \over \delta A_\rho^a(\gamma_u(v))}
 U_{\gamma_u}(s,v) T^a U_{\gamma_u}(v,t).\nonumber\\
\label{martin}
\eear  Using (\ref{derivada}) this can be written as:
\bear && {\hskip-0.8cm} {d\over du} \langle U_{\gamma_u}(s,t) \rangle= 
{4\pi i \over k}\int [DA]
\ex^{i S(A)} \int_s^t dv\, \epsilon_{\mu\nu\rho}
\gamma_u^{'\mu}(v) \dot\gamma_u^\nu(v) \nonumber\\ && \Big[\int_s^v dw
\dot\gamma_u^\rho(w)
\delta^{(3)}(\gamma_u(v),\gamma_u(w))
  U_{\gamma_u}(s,w) T^a U_{\gamma_u}(w,v) T^a U_{\gamma_u}(v,t)
\nonumber\\ && +\int_v^t dw \dot\gamma_u^\rho(w)
\delta^{(3)}(\gamma_u(v),\gamma_u(w)) U_{\gamma_u}(s,v) T^a
U_{\gamma_u}(v,w) T^a U_{\gamma_u}(w,t)\Big].
\nonumber\\
\label{martinez}
\eear Due to the presence of the delta function in this expression, 
contributions come from the solutions to the equation
\beq
\gamma_u(v)=\gamma_u(w).
\label{ruben}
\eeq This equation has two types of solutions. The first type is the
one-dimensional set of solutions $v=w$ for each value of $u$. The second type
is made out of the solutions for which $v\neq w$. Notice that if there are
not self-intersections the set of second-type solutions is empty.  Since
(\ref{martinez}) is only non-vanishing for $v$ in a neighborhood of
$s_1$, we have only one second-type solution: $u=0$, $v=s_1$ and $w=t_1$.
The contribution comes from the second $w$-integral in (\ref{martinez}).

Contributions of the first type are related to framing. Recall that in
Chern-Simons gauge theory one does not have topological invariants unless
a framing  is introduced. If there is not framing there is a contribution
proportional to the Gauss self-linking number which is metric dependent \cite{gmm}.
As argued in \cite{vande}, the variation of this contribution under a
deformation of the path leads in fact to (\ref{martinez}) for
contributions of the first type. In the case one introduces a framing,
contributions of the first type vanish because the
paths entering the delta function (the path itself and its companion)
never get values at coincident points in three-space for
$v=w$. Thus, if one considers framed knots, contributions of the first
type can be ignored. In the case of framed knots there is always a
contribution of the second type if the path has a self-intersection.
Generically it will occur for values of $u$, $v$ and $w$ slightly
displaced from from the previously assigned values at the intersection
point. In the limit of zero width one recovers the initial values. The
contribution is framing independent and one can therefore continue the
analysis of  (\ref{martinez}) without taking into account the framing and
ignoring the contributions of first type. See \cite{brug} for a detailed
analysis of the contributions to (\ref{martinez}) of the first type.

For the family of paths under consideration there is only one contribution
of the second type in the integral (\ref{martinez}). This occurs at 
$u=0$, $v=s_1$ and $w=t_1$. In a neighborhood of this solution,  the
delta-function can be written as:
\beq
\delta^{(3)}(\gamma_u(v),\gamma_u(w))= {1\over|\Delta(0,s_1,t_1)|}\delta(u)
\delta(v-s_1)\delta(w-t_1),
\label{ladelta}
\eeq where,
\beq
\Delta(u,v,w) = \epsilon_{\mu\nu\rho}\gamma_u^{'\mu}(v)
\dot\gamma_u^\nu(v)
 \dot\gamma_u^\rho(w) .
\label{ladeltaza}
\eeq Plugging this expression into the derivative (\ref{martinez}) one
finds that, as expected, it is proportional to a delta function in the
variable
$u$:
\beq {d\over du} \langle U_{\gamma_u}(s,t) \rangle=  {4\pi i \over
k}\delta(u) 
  \int [DA]
\ex^{i S(A)}  U_{\gamma_u}(s,s_1) T^a U_{\gamma_u}(s_1,t_1) T^a
U_{\gamma_u}(t_1,t).
\label{martinazo}
\eeq  In obtaining this expression from (\ref{martinez}) one has to take
into account that for the selected values of $u$, $v$ and $w$ one has
$\Delta(0,s_1,t_1)>0$. Integrating in the variable
$u$ in a region which encloses
$u=0$ one finds the difference between $\langle U(s,t)_+\rangle$, the
Wilson line with no intersection and an overcrossing at $P$ and $\langle
U(s,t)_-\rangle$, the Wilson line with no intersections and an
undercrossing at $P$:
\bear  && {\hskip-1cm}
\Big\langle U(s,t)_+\Big\rangle  - \Big\langle U(s,t)_-\Big\rangle=
\int_{-\eta}^\eta du  {d\over du} \Big\langle U_{\gamma_u}(s,t)
\Big\rangle 
\nonumber\\ && =
 {4\pi i \over k}
\int [DA]
\ex^{i S(A)}  U_{\gamma_0}(s,s_1) T^a U_{\gamma_0}(s_1,t_1) T^a
U_{\gamma_0}(t_1,t),  
\label{martinezdos}
\eear where $\eta$ is some  positive small  real number.

From the previous expression one can read very simply the form of the
operator associated to a Wilson line with a double point at a point
$P\in M$ which corresponds to the values $v=s_1$ and $v=t_1$ of its path
parameter. The rule is very simple: the original Wilson line is splited
into three sections. The operator is made out of the ordered product of
the corresponding three Wilson lines with insertions of group generators
at each joint. The operator obtained in (\ref{martinezdos}) is gauge
covariant. Recall that under a gauge transformation,
\beq A_\mu \rightarrow \Lambda^{-1} (\partial_\mu+A_\mu) \Lambda,
\label{transformacion}
\eeq a Wilson line $U_\gamma(s,t)$ transforms as,
\beq U_\gamma(s,t) \rightarrow \Lambda^{-1}(\gamma(s)) U_\gamma(s,t) 
\Lambda(\gamma(t)).
\label{transformacionw}
\eeq Thus, the operator inserted in (\ref{martinezdos}) transforms as:
\bear && {\hskip-1cm} U_{\gamma}(s,s_1) T^a U_{\gamma}(s_1,t_1) T^a
U_{\gamma}(t_1,t) \rightarrow \nonumber\\ &&
\Lambda^{-1}(\gamma(s)) U_{\gamma}(s,s_1) \Gamma^a(\gamma(s_1))
U_{\gamma}(s_1,t_1)
\Gamma^a(\gamma(t_1)) U_{\gamma}(t_1,t)\Lambda(\gamma(t)),
\nonumber \\
\label{mastras}
\eear where,
\beq
\Gamma^a(x) = \Lambda^{-1}(x) T^a \Lambda(x) =
\lambda^{ab}(x) T^b.
\label{conju}
\eeq In  this last equation we have used the fact that the set of group
generators is invariant under conjugation. The quantities
$\lambda^{ab}$ entering (\ref{conju}) satisfy:
\beq
\lambda^{ab}(x) \lambda^{ac}(x) = \delta^{bc}.
\label{alicante}
\eeq Using this property and the fact that $\gamma(s_1)=\gamma(t_1)$ one
finds that, indeed, the operators entering (\ref{martinezdos}) are gauge
covariant:
\bear && {\hskip-1cm} U_{\gamma}(s,s_1) T^a U_{\gamma}(s_1,t_1) T^a
U_{\gamma}(t_1,t) \rightarrow \nonumber\\ &&
\Lambda^{-1}(\gamma(s)) U_{\gamma}(s,s_1) T^a U_{\gamma}(s_1,t_1) T^a
U_{\gamma}(t_1,t)\Lambda(\gamma(t)).
\label{remastras}
\eear

The result (\ref{martinezdos}) can be used to obtain the operator
associated to a singular knot with one double point. Let us denote by
$\psi_+$ the vacuum expectation value corresponding to a Wilson loop when
the double point at $P\in M$ is solved by an overcrossing, and
$\psi_-$ when it is solved by an undercrossing. The difference 
$\psi_+-\psi_-$ will be given by the trace of (\ref{martinezdos}) for a
closed path, \ie\ a  path $\gamma$ where its end points are identified.
One easily finds:
\beq 
\psi_+-\psi_-= 
 2x {\Bigg\langle}
\tr\Big[T^a
 U_{\gamma}(s_1,t_1) T^a U_{\gamma}(t_1,s_1)\Big] {\Bigg\rangle},  
\label{primero}
\eeq where, again, the subindex in the path $\gamma$ has been suppressed.
In this equation we have made use of (\ref{laequis}). The gauge invariance
of the operator entering this expression follows trivially from
(\ref{remastras}).

As a consistency check of this equation, let us compute its lowest order
contribution for a singular knot $K^1$ with one double point. We will
denote by $K_+$ and $K_-$ the non-singular knots which correspond to the
resolution of the double point by an overcrossing and an undercrossing,
respectively. If one chooses a framing in which the linking number of a
knot and its companion knot coincide with the writhe (for some projection
of the knot)  one has that at the lowest non-trivial order the vacuum
expectation value of its associated Wilson loop is proportional to the
writhe. Indeed, one finds, for example, for $\psi_+$:
\beq
\psi_+ = {\rm dim} R\,\Big(1+x\, C_2(R) \sum_{i}\epsilon_i\Big),
\label{andrea}
\eeq where the sum is over all the crossings $i$, and $\epsilon_i$ is the
sign of the crossing $i$. In (\ref{andrea}) $C_2(R)=\tr(T^aT^a)/{\rm
dim}R$ and $R$ is the representation carried by the knot. A similar
equation holds for the lowest order of $\psi_-$. The difference
$\psi_+-\psi_-$ can be computed very simple because all the terms of their
sums are the same for both knots except the one corresponding to the
double point. The contribution from all but one, the double point, cancel.
The contributions from the double point add and one ends with
\beq
\psi_+ - \psi_- =  2x\,{\rm dim} R \, C_2(R) =2x\,\tr\big( T^a T^a\big).
\label{andreados}
\eeq This is precisely what one obtains from (\ref{primero}) at lowest
order.

Our next task is to analyze differences corresponding to two singular
knots with one double point to obtain the form of the operators for
singular knots with two double points. The strategy is the following. Let
us consider a singular knot with two double points. One of them is located
at a point $P_1\in M$ and the corresponding values of the knot parameter
are $v=s_1$ and $v=t_1$, with $s_1<t_1$. The second one is located at
$P_2\in M$ for $v=s_2$ and $v=t_2$, with
$s_2<t_2$. The resolution of the second double point into an overcrossing
and an undercrossing involves two singular knots with one double point
whose invariants will be denoted by
$\psi^1_+$ and $\psi^1_-$, respectively. The singular knot invariants
associated to $\psi^1_\pm$ can be written as the vacuum expectation values:
\beq
\psi^1_\pm = 2x\, \Bigg\langle
\tr\Big[T^a U_{\gamma_\pm}(s_1,t_1) T^a U_{\gamma_\pm}(t_1,s_1)\Big]
\Bigg\rangle,
\label{neha}
\eeq where $\gamma_\pm$ denotes the paths associated to the singular knots
corresponding to the two resolutions. In order to find out the operator
corresponding to a singular knot with two double points we must consider a
family of singular knots parametrized by $u$ such that for
$u>0$ ($u<0$) the singular knot is equivalent under rigid vertex isotopy
to $\psi^1_+$ ($\psi^1_-$), and for $u=0$ it possesses a second double
point at $P_2$. Making use of this family, the difference $\psi^1_+-
\psi^1_-$ can be written as:
\beq
\psi^1_+- \psi^1_- = \, \int_{-\eta}^\eta du {d\over du}
\psi^1_{\gamma_u},
\label{dinas}
\eeq where,
\beq
 \psi^1_{\gamma_u} = 2x \Bigg\langle
\tr\Big[T^a U_{\gamma_u}(s_1,t_1) T^a U_{\gamma_u}(t_1,s_1)\Big]
\Bigg\rangle.
\label{moeko}
\eeq

\begin{figure}
\centerline{\hskip.4in \epsffile{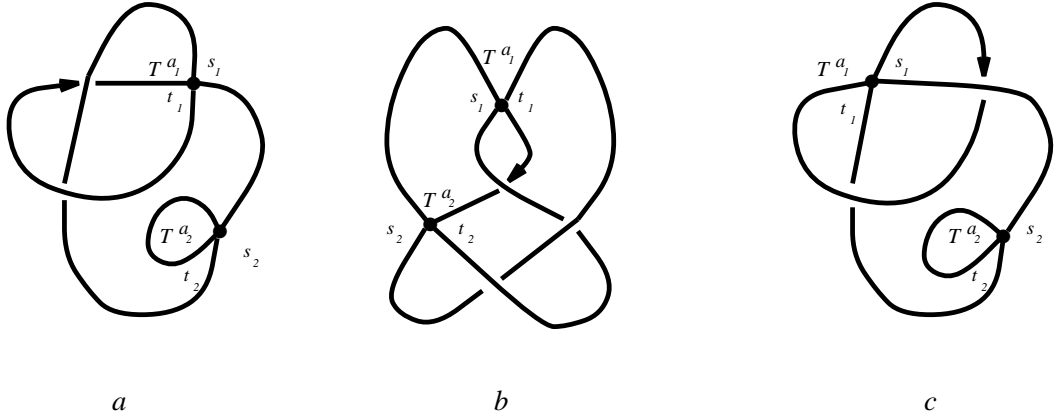}}
\caption{Examples of the three types of contributions.}
\label{example}
\end{figure}

To analyze the derivative of $\psi^1_{\gamma_u}$ respect to $u$ let us
assume that $s_1<s_2<t_1$. As it will become clear from the analysis, the
other case, $t_1<s_2$, can be treated similarly. Using  (\ref{variacion})
and (\ref{fieldequation})    and integrating by parts in connection space
one finds:
\bear && {d\over du} \psi^1_{\gamma_u}=  (2x)^2\int [DA]
\ex^{i S(A)} \nonumber\\ && {\hskip-0.8cm} 
\int_s^t dv\, \epsilon_{\mu\nu\rho}\gamma_u^{'\mu}(v)\dot\gamma_u^\nu(v) 
 {\delta \over \delta A_\rho^b(\gamma_u(v))} \tr\Big[ T^a
U_{\gamma_u}(s_1,v) T^b U_{\gamma_u}(v,t_1) T^a
U_{\gamma_u}(t_1,s_1)\Big],\nonumber\\
\label{emily}
\eear which, after using (\ref{derivada}) can be written as:
\bear && {d\over du} \psi^1_{\gamma_u}=   (2x)^2\int [DA]
\ex^{(i S(A))} 
\int_s^t dv\, \epsilon_{\mu\nu\rho} \gamma_u^{'\mu}(v)\dot\gamma_u^\nu(v)
\nonumber\\ && {\hskip-0.8cm} 
\Bigg[ 
\int_{s_1}^v dw\,
\dot\gamma_u(w)\delta^{(3)}(\gamma_u(v),\gamma_u(w)) \tr\Big[ T^a
U_{\gamma_u}(s_1,w) T^b U_{\gamma_u}(w,v) T^b U_{\gamma_u}(v,t_1) T^a
U_{\gamma_u}(t_1,s_1)\Big],\nonumber\\ && {\hskip-1.1cm}  +\int_{v}^{t_1}
dw\,
\dot\gamma_u(w)\delta^{(3)}(\gamma_u(v),\gamma_u(w)) \tr\Big[ T^a
U_{\gamma_u}(s_1,v) T^b U_{\gamma_u}(v,w) T^b U_{\gamma_u}(w,t_1) T^a
U_{\gamma_u}(t_1,s_1)\Big],\nonumber\\ && {\hskip-1.1cm} 
+\int_{t_1}^{s_1} dw\,
\dot\gamma_u(w)\delta^{(3)}(\gamma_u(v),\gamma_u(w)) \tr\Big[ T^a
U_{\gamma_u}(s_1,v) T^b U_{\gamma_u}(v,t_1) T^a U_{\gamma_u}(t_1,w) T^b
U_{\gamma_u}(w,s_1)\Big]\Bigg].\nonumber\\
\label{laureen}
\eear As in the previous case we must now study the solutions to equation
(\ref{ruben}). Certainly, there are solutions of the first type. These, as
argued before, can be ignored. In the region where
$\gamma_u^{'\nu}(v)\neq 0$ we have only one solution of the second type:
$v=s_2$ and $w=t_2$. The only term of the three-dimensional delta function
in (\ref{laureen}) which contributes can be written as,
\beq {1\over |\Delta(0,s_2,t_2)|}\delta(u)\delta(v-s_2)\delta(w-t_2),
\label{masdelta}
\eeq where $\Delta(u,v,w)$ is given in (\ref{ladeltaza}). Only one of the
three integrals in (\ref{laureen}) contribute. One must distinguish two
cases: $t_2 < t_1$ and $t_2 > t_1$. In the first case one ends with
(recall that we are considering the situation $s_2<t_1$):
\beq  (2x)^2\,\delta(u)\int[DA]\ex^{iS(A)}
\tr\Big[T^a U_{\gamma_u}(s_1,s_2) T^b U_{\gamma_u}(s_2,t_2) T^b
U_{\gamma_u}(t_2,t_1)T^a U_{\gamma_u}(t_1,s_1)\Big],
\label{once}
\eeq while in the second case,
\beq  (2x)^2\,\delta(u)\int[DA]\ex^{iS(A)}
\tr\Big[T^a U_{\gamma_u}(s_1,s_2) T^b U_{\gamma_u}(s_2,t_1) T^a
U_{\gamma_u}(t_1,t_2)T^b U_{\gamma_u}(t_2,s_1)\Big].
\label{doce}
\eeq It is also clear that in the situation $t_1<s_2$ we would have
obtained:
\beq  (2x)^2\,\delta(u)\int[DA]\ex^{iS(A)}
\tr\Big[T^a U_{\gamma_u}(s_1,t_1) T^a U_{\gamma_u}(t_1,s_2) T^b
U_{\gamma_u}(s_2,t_2)T^b U_{\gamma_u}(t_2,s_1)\Big].
\label{trece}
\eeq Using (\ref{dinas}) we finally obtain the final expression for 
$\psi^1_+- \psi^1_-$. There are three cases depending on the values  of
$s_2$ and $t_2$:
\bear  && {\hskip-1cm} (2x)^2 \Bigg\langle\tr\Big[ T^a U(s_1,s_2) T^b
U(s_2,t_2) T^b U(t_2,t_1)T^a U(t_1,s_1)\Big]\Bigg\rangle, {\hskip0.2cm}
s_1 < s_2 < t_2 < t_1, \nonumber\\ && {\hskip-1cm} (2x)^2
\Bigg\langle\tr\Big[ T^a U(s_1,s_2) T^b U(s_2,t_1) T^a U(t_1,t_2)T^b
U(t_2,s_1)\Big]\Bigg\rangle, {\hskip0.2cm} s_1 < s_2 < t_1 < t_2,
\nonumber\\ && {\hskip-1cm} (2x)^2 \Bigg\langle\tr\Big[ T^a U(s_1,t_1) T^a
U(t_1,s_2) T^b U(s_2,t_2)T^b U(t_2,s_1)\Big]\Bigg\rangle, {\hskip0.2cm}
s_1 < t_1 < s_2 < t_2, \nonumber\\
\label{greg}
\eear where we have suppressed the subindex of the Wilson line which
denoted the path. The three different situations  have been depicted in
fig. \ref{example}a, \ref{example}b and \ref{example}c.  Notice that the
rule to write down the corresponding operators is very simple. One labels
the double points, chooses one of them as base point, and  travels along
the singular knot inserting a generator at each double point, and a Wilson
line for each section between double points.

The results (\ref{greg}) can written in a compact form introducing the
triples
$\tau_i=\{s_i,t_i,T^{a_i}\}$ for $i=1,2$ and the set
$\{s_1,t_1,s_2,t_2\}$.  Ordering the elements of this set into 
$\{w_1,w_2,w_3,w_4\}$, $w_i < w_{i+1}$, and constructing the map
$\phi(w_i)$ which assigns to $w_i$ the index of the group generator in the
triple to which $w_i$ belongs, one can write the three cases as:
\bear && {\hskip-1cm}
\psi^1_+- \psi^1_- = \nonumber\\ && {\hskip-1cm} (2x)^2
\Bigg\langle\tr\Big[ T^{\phi(w_1)} U(w_1,w_2) T^{\phi(w_2)} U(w_2,w_3)
T^{\phi(w_3)} U(w_3,w_4) T^{\phi(w_4)} U(w_4,w_1) 
\Big]\Bigg\rangle \nonumber\\
\label{laura}
\eear where we have dropped the subindex $\gamma$ of $U$ which specified
the path.  This is precisely the operator (\ref{operador}) for $n=2$.

The picture that emerges for the general situation is the following. Given
an operator for a singular knot with $n-1$ double points, the operator
corresponding to a singular knot with one more double point will be the
result of inserting an operator of the form $UT^aU$ at the two sections
of the singular knot involved in the additional double point. Let us
assume that we have a singular knot with $n$ double points and that we
have constructed the operators for all singular knots with $n-1$ double
points. The $n^{\rm th}$ double point will generally involve two sections,
the one corresponding to the open interval $(w_i, w_{i+1})$ in parameter
space, and the one corresponding to $(w_j, w_{j+1})$, $w_i < w_j$. It
might happen that the resolution of the  $n^{\rm th}$ double point
involves only one section, as it is the case of the second double point
shown in figs.
\ref{example}a and \ref{example}c. The analysis that follows holds also for that case. The new
double point is located at the points $s_n$ and $t_n$,
$s_n<t_n$,  in parameter space. Of course, $w_i < s_n < w_{i+1}$ and $w_j
< t_n < w_{j+1}$. The operators associated to the singular knots with
$n-1$ double points in the resolution of the new double point will have
the form:
\beq (2x)^{n-1} \tr\Big[\cdots U(w_i,w_{i+1})\cdots
U(w_j,w_{j+1})\cdots\Big].
\label{movenpick}
\eeq We need to compute the difference of the vacuum expectation values
associated to these operators, $\psi^{n-1}_+ - \psi^{n-1}_-$, which, as
before, will be computed studying the first derivative respect to $u$ 
of the vacuum expectation value of the members of a family parametrized to
$u$ which continuously connects one resolution to the other and that for
$u=0$ corresponds to the singular knot with $n$ double points. We have:
\beq
\psi^{n-1}_+ - \psi^{n-1}_- =
\int_{-\eta}^\eta du \, {d\over du} \psi^{n-1}_{\gamma_u},
\label{aristocats}
\eeq where
\beq
\psi^{n-1}_{\gamma_u}= (2x)^{n-1} \Bigg\langle\tr\Big[\cdots
U_{\gamma_u}(w_i,w_{i+1})\cdots
U_{\gamma_u}(w_j,w_{j+1})\cdots\Big]\Bigg\rangle.
\label{gatos}
\eeq As in the previous cases the loop $\gamma_u$ differs from $\gamma_0$
only in a region near the point $s_n$ in parameter space. Only in a small
neighborhood of $s_n$ the derivative $\gamma_u^{'}$ does not vanish. To
compute  the derivative of (\ref{gatos}) we must proceed as before. The
use of (\ref{variacion}) leads to a replacement of 
$U_{\gamma_u}(w_i,w_{i+1})$ by
$U_{\gamma_u}(w_i,v)T^{a_n}U_{\gamma_u}(v,w_{i+1})$:
\bear && {\hskip-1cm} {d\over du} \psi^{n-1}_{\gamma_u}=  (2x)^n\int [DA]
\ex^{i S(A)}  
\int_s^t dv\, \epsilon_{\mu\nu\rho}\gamma_u^{'\mu}(v)\dot\gamma_u^\nu(v) 
\nonumber\\ &&  {\delta \over \delta A_\rho^{a_n}(\gamma_u(v))} \tr\Big[
\cdots U_{\gamma_u}(w_i,v)T^{a_n}U_{\gamma_u}(v,w_{i+1}) \cdots
U_{\gamma_u}(w_j,w_{j+1})\cdots
\Big].\nonumber\\
\label{emilydos}
\eear The functional derivative respect to $A_\rho^{a_n}(\gamma_u(v))$
acts on each of the Wilson lines present in (\ref{emilydos}). However, due
to the presence of the delta function, upon computing the functional
derivative (see (\ref{derivada})), the only contribution which survives is
the one coming from the Wilson line
$U_{\gamma_u}(w_j,w_{j+1})$. One obtains in this way an insertion of
$U_{\gamma_u}(w_j,w)T^{a_n}U_{\gamma_u}(w,w_{j+1})$ in the place of
$U_{\gamma_u}(w_j,w_{j+1})$ and then, after integrating in $v$ and $w$,
one finds:
\bear && {\hskip-1cm} {d\over du} \psi^1_{\gamma_u}=(2x)^n\,\delta(u)
\Bigg\langle
\tr\Big[\cdots   U_{\gamma_u}(w_i,s_n)T^{a_n}U_{\gamma_u}(s_n,w_{i+1})
\cdots
\nonumber\\&& {\hskip5cm}
\cdots U_{\gamma_u}(w_j,t_n)T^{a_n}U_{\gamma_u}(t_n,w_{j+1})\cdots
\Big]\Bigg\rangle \nonumber\\
\label{disney}
\eear which, after using (\ref{aristocats}), leads to:
\bear && {\hskip-0.7cm}
\psi^{n-1}_+ - \psi^{n-1}_- = \nonumber\\ &&  (2x)^n \Bigg\langle
\tr\Big[\cdots   U(w_i,s_n)T^{a_n}U(s_n,w_{i+1})
\cdots U(w_j,t_n)T^{a_n}U(t_n,w_{j+1})\cdots
\Big]\Bigg\rangle,\nonumber\\
\label{dalmatians}
\eear where we have suppressed the subindices of the Wilson lines which
denoted the loop. It is clear from the analysis that if in the resolution
of the double point there is only one section involved the result is
\bear && {\hskip-0.7cm}
\psi^{n-1}_+ - \psi^{n-1}_- = \nonumber\\ &&  (2x)^n \Bigg\langle
\tr\Big[\cdots   U(w_i,s_n)T^{a_n}U(s_n,t_n)T^{a_n}
\cdots U(t_n,w_{i+1})\cdots
\Big]\Bigg\rangle,\nonumber\\
\label{chuchos}
\eear Both situations fit with the expression announced in 
(\ref{operador}). 

The gauge invariance of the operators (\ref{operador}) follows from
(\ref{transformacionw}), (\ref{conju}) and (\ref{alicante}). Indeed, one
has that under the gauge transformation (\ref{transformacion}) the
operator (\ref{operador}) becomes,
\bear && {\hskip-1cm} (2x)^n \tr \Big[\Gamma^{\phi(w_1)}(\gamma(w_1))
U_\gamma(w_1,w_2) \Gamma^{\phi(w_2)}(\gamma(w_2)) U_\gamma(w_2,w_3)
\Gamma^{\phi(w_3)} (\gamma(w_3))\cdots \nonumber\\ && {\hskip3cm} \cdots
U_\gamma(w_{2n-1},w_{2n}) \Gamma^{\phi(w_{2n})}(\gamma(w_{2n}))
U_\gamma(w_{2n},w_{1}) \Big],
\nonumber\\
\label{gerona}
\eear where we have made use of (\ref{transformacion}) and
(\ref{transformacionw}). Taking into account (\ref{conju}) this expression
can be written as:
\bear && {\hskip-1cm} (2x)^n \lambda^{\phi(w_1) b_1}(\gamma(w_1))
\cdots \lambda^{\phi(w_{2n})) b_{2n}}(\gamma(w_{2n}))
\nonumber\\ && \tr \Big[T^{b_1} U_\gamma(w_1,w_2) T^{b_2}
U_\gamma(w_2,w_3) T^{b_3}  \cdots
          U_\gamma(w_{2n-1},w_{2n}) T^{b_{2n}} U_\gamma(w_{2n},w_{1})
\Big].
\nonumber\\
\label{barna}
\eear The indices $\phi(w_1),\dots,\phi(w_{2n})$ are paired and precisely
when $\phi(w_i)=\phi(w_j)$ one has $\gamma(w_i)=\gamma(w_j)$. We can then
apply (\ref{alicante}) $n$ times, obtaining (\ref{operador}) and therefore
proving its gauge invariance.
\vskip1cm In this section, we have shown the following:
\vskip0.5cm

Let $K^n$ be a singular knot with $n$ double points, and let us assign to
each double point $i$ a triple $\tau_i=\{s_i,t_i,T^{a_i}\}$ where $s_i$ and
$t_i$, $s_i<t_i$, are the values of the $K^n$-parameter at the double
point, and $T^{a_i}$ is a group generator. The gauge invariant operator
associated to the singular knot $K^n$ is:
\bear && {\hskip-2cm} (2x)^n \tr \Big[T^{\phi(w_1)} U(w_1,w_2)
T^{\phi(w_2)} U(w_2,w_3) T^{\phi(w_3)}\cdots \nonumber\\ && {\hskip3cm}
\cdots
U(w_{2n-1},w_{2n}) T^{\phi(w_{2n})} U(w_{2n},w_{1}) \Big],
\nonumber\\
\label{operadordos}
\eear where $\{w_i;i=1,\dots,2n\}$, $w_i<w_{i+1}$, is the set which
results of ordering the values $s_i$ and $t_i$, for $i=1,\dots,n$, and
$\phi$ is a map which assigns to each $w_i$ the group generator in the
triple to which it belongs.

\vfill
\eject

\section{Immediate applications}
\setcounter{equation}{0}

In this section we will describe some immediate applications of the
operators (\ref{operador}). First we will show that the theorem by Birman
and Lin \cite{bilin} proving that the $n$-order coefficient of the
expansion of a (quantum group or Chern-Simons) polynomial knot invariant
(after replacing its variable by
$\ex^x$) is a Vassiliev invariant of order $n$ easily follows from the
form of the operators (\ref{operador}). Then, we will show that the form
of the top row of actuality tables based on semi-simple Lie algebras
provided by Bar-Natan \cite{barnatan} (Lie-algebra based weight systems)
is also a direct consequence of the form of these operators. Actually, we
will argue that our construction constitutes a gauge independent proof of
the fact that Bar-Natan's weight systems can be integrated to Vassiliev
invariants for non-singular knots. 

\subsection{Birman and Lin theorem}

Some years ago Birman and Lin proved that if in any polynomial knot
invariant with variable $t$ one substitutes $t\rightarrow \ex^x$, and
expand in powers of $x$, the coefficient of the power $x^n$ is a Vassiliev
invariant of order $n$. A Vassiliev invariant of order $n$ vanishes for
all singular knots with $n+1$ crossings. In the language of Chern-Simons
gauge theory the theorem by Birman and Lin can be rephrased simply stating
that the $n$-order term of the perturbative series expansion of the vacuum
expectation value of a the Wilson loop associated to a given knot is a
Vassiliev invariant of order $n$. We will now prove that this is so with
the help of the operators (\ref{operador}).

The vacuum expectation value of the operators (\ref{operador}) provide an
invariant for a singular knot with $n$ double points. This singular-knot
invariant can be expressed as a signed sum of $2^n$ invariants for
non-singular knots. These $2^n$ invariants are the perturbative series
expansion of the vacuum expectation value of the corresponding Wilson
loops. To show that the coefficient of order $n$ of these vacuum
expectation values is a Vassiliev invariant or order $n$ is equivalent to
prove that all the terms of order less than $n$ vanish in the signed sum.
But the signed sum is precisely the vacuum expectation value of the
operator  (\ref{operador}). Since the perturbative series expansion of
these operators starts at order $n$, the proof of the theorem by Birman
and Lin follows.

\subsection{Bar-Natan's weight systems}

In \cite{barnatan}  weight systems were introduced as elements of the top
row of actuality tables for Vassiliev invariants. Using the Knotsevich
integral, it was proved there that weight systems can be integrated to
construct Vassiliev invariants. In this subsection we show that the
operators (\ref{operador}) at lowest order are  weight systems for the case of semi-simple Lie algebras. As
we will argue below, our construction constitutes a gauge independent
proof of the integrability of these weight systems to Vassiliev
invariants for non-singular knots. We call it gauge independent because
from the point of view of Chern-Simons gauge theory the original proof in
\cite{barnatan} can be regarded as performed in a given gauge, the
light-cone gauge. As it was shown in \cite{cata,lcone}, the Kontsevich integral
appears in the context of Chern-Simons gauge theory when the theory is
analyzed  from a perturbative point of view in the light-cone gauge. In
the analysis presented in this paper we have not been forced to make a
gauge choice and therefore it is gauge independent. 

Let us consider the vacuum expectation values of the operators
(\ref{operador}) at order zero. Their expression is simply obtained
setting $U=1$ in all the Wilson lines. Let us consider a singular knot
$K^n$  with $n$ double points, 
$s_i$, $t_i$, with $s_i<t_i$, for $i=1,\dots,n$, in parameter space. As
usual, with these data we construct the triples: $\tau_i=\{s_i,t_i,
T^{a_i}\}$. At lowest order in perturbation theory the operators
(\ref{operador}) become the group factors,
\beq v_n (K^n) = (2x)^n \tr \Big[T^{\phi(w_1)} T^{\phi(w_2)}
T^{\phi(w_3)}\cdots T^{\phi(w_{2n})} \Big],
\nonumber\\
\label{wsystem}
\eeq where the set $\{w_1,w_2,\dots,w_n\}$, with $w_j < w_{j+1}$, is
obtained by ordering the set $\{s_i,t_i; i=1,\dots,n\}$, and $\phi$ is the
induced map which assigns to each $w_j$ the index of the group generator
in the triple to which $w_j$ belongs. The indices entering
(\ref{wsystem}) are paired. This allows to associate to each operator
(\ref{wsystem}) a Feynman-like diagram in which the $2n$ points are
distributed in a circle and the ones which possess the same value of
$\phi$ are joined by a line. In other words one can use simply the
group-theoretical Feynman rules emerging from Chern-Simons gauge theory
(see, for example, \cite{lcone}). The resulting diagrams are the
configuration diagrams for singular knots (as the one shown in fig.
\ref{confi}) and correspond to Bar-Natan's chord diagrams with $n$
chords \cite{barnatan}. 

It was proved in \cite{barnatan} that if weight systems are regarded as
Vassiliev invariants for singular knots they can be integrated to
Vassiliev invariants for non-singular knots. Let us consider an element of
a weight system with $n$ chords, $B^n$, and any of its associated singular
knots with $n$ double points, $K^n$. What was proved in
\cite{barnatan} is that  $B^n$  can be written as a signed sum of the
$2^n$ Vassiliev invariants of non-singular knots which are generated when
one performs the resolutions of the double points of $K^n$. The proof was
done making use of the Kontsevich integral for Vassiliev invariants of
non-singular knots. It was shown that the associated signed sum of
Kontsevich integrals starts with a chord diagram of order $n$. But this is
what has been done in this paper. Here, starting with the vacuum
expectation values of Wilson loops associated to  non-singular knots we
have proved that the signed sum entering the invariant associated to the
singular knot $K^n$ starts at order $n$, with the group factor
(\ref{wsystem}). This group factor plays the role of the order-$n$ chord
diagram appearing in Bar-Natan's approach.

As argued above, our work contains a gauge independent proof of
Bar-Natan's theorem for the case of weight systems coming from
semi-simple Lie algebras. Being a gauge theory, Chern-Simons gauge theory
possesses a variety of pictures depending on the gauge in which it is
studied. It was shown in
\cite{lcone} that in the light-cone gauge the picture which appears
involves the Knotsevich integral. Thus, in a field-theory language,
Bar-Natan's theorem was proved in the light-cone gauge. What is contained
in this paper is a gauge independent proof of that theorem. Of course,
the theorem has to hold in other gauges and in each gauge it can teach us
something new. For example, it would be interesting to study it in a
covariant gauge.

The group factors (\ref{wsystem}) are Vassiliev invariants of order $n$
for singular knots with $n$ double points. Of course, not all them are
independent. Due to the Lie-algebra relations satisfied by the group
generators there are linear relations between them.  One
important open problem in the theory of Vassiliev invariants is how many
independent group factors (\ref{wsystem}) are at each order. The
answer to this question is only known for low values of $n$
\cite{barnatan}. Once a choice of independent group factors is made one
can build the top row of the actuality table for Vassiliev invariants of
order $n$. 

\vfill
\eject

\section{Kontsevich integral for singular knots}
\setcounter{equation}{0}

In this section we will analyze the perturbative series expansion
corresponding to the vacuum expectation value of the operators
(\ref{operador}). We will carry out the analysis in the light-cone gauge.
As a result, we will obtain a generalization of the Kontsevich integral for
Vassiliev invariants which is also valid  in the case of singular knots.

Chern-Simons gauge theory in the light-cone gauge was first studied in
\cite{king} where the vacuum expectation value of Wilson loops were
analyzed from a non-perturbative point of view. In that paper a close
relation between the vacuum expectation value of a Wilson line and the
Knizhnik-Zamolodchikov  \cite{kzeqs} equations was pointed out. The loop
structure of Chern-Simons gauge theory in the light-cone gauge has been
studied in \cite{leibbrandt,sorella}. Recently, the analysis of the
perturbative series expansion of the vacuum expectation values of Wilson
lines has been carried  out \cite{lcone}, showing that it contains the
Kontsevich integral for Vassiliev invariants of framed knots. In this
section we will extend that analysis to the case in which the vacuum
expectation values of the operators (\ref{operador}) associated to
singular knots are considered. 

To achieve our goal we will first review some basic facts about the
light-cone gauge. We refer the reader to \cite{king,lcone} for more
details. The light-cone gauge condition has the form:
\beq n^\mu A_\mu = 0,
\label{condicion}
\eeq where $n^\mu$ is the light-like vector $(0,1,-1)$. Introducing the 
light-cone coordinates,
\beq x^+ = x^1 + x^2,   \,\,\, \,\,\, x^- = x^1-x^2,
\label{coor}
\eeq  and the corresponding light-cone components for the connection,
\beq A_+ = A_1 + A_2,   \,\,\, \,\,\, A_- = A_1-A_2,
\label{lcgc}
\eeq  the gauge-fixing condition (\ref{condicion}) takes the form $A_- =
0$.  The Fadeev-Popov ghost fields corresponding to the gauge fixing
(\ref{condicion}) decouple in the quantum action. The relevant part of
this action for our purposes takes the form:
\beq  S_q(A)={k\over 4\pi}\int_{M} d^3 x \, \big( A^a_+\partial_- A^a_0 -
A_0^a\partial_- A_+^a\big).
\label{qaction}
\eeq
In this section we will consider the case $M=\RR^3$.

As explained in \cite{lcone}, to consider (\ref{qaction}) as the relevant
quantum action to compute vacuum expectation values of gauge invariant
operators is an oversimplification. The perturbation-theory formulation of
gauge theories in non-covariant gauges is a very delicate issue
 \cite{leibrew}. On the one hand, propagators contain unphysical poles
which have to be handled introducing some prescription; on the other hand,
there is a residual gauge invariance. In the light-cone gauge the gauge
fixing condition (\ref{condicion}) is preserved under gauge transformations 
(\ref{transformacion}) such that $\Lambda$ depends only on
$x^0$ and $x^+$. Both problems are related. It is a widespread belief
that an adequate choice of prescription to avoid the unphysical poles
solves the problem of the residual gauge invariance. In
\cite{king,leibbrandt,lcone} the light-cone gauge has been analyzed using
the Mandelstan-Leibbrandt prescription. We think that this prescription is
not enough to solve the problem of the residual gauge invariance. As
discussed in \cite{lcone}, the action (\ref{qaction}) together with the
Mandelstan-Leibbrandt prescription to circumvent the unphysical poles does
not lead to a topological invariant expression for the vacuum expectation
value of a Wilson loop. One needs to introduce a correction to the
perturbative result to obtain the correct answer. We believe that this
correction will be obtained after solving the problem related to the
residual gauge invariance. The issue of the residual gauge invariance has
been treated in \cite{vande} for the temporal gauge. For the purposes of
this paper it is enough to consider (\ref{qaction}) as the relevant
quantum action of the theory. Of course, doing this we do not expect to
obtain the correction term. However, we will certainly construct the
generalization of the Kontsevich integral for the case of singular knots.

Another important issue which was discussed in \cite{lcone} is the
contribution from higher loops. This problem has been addressed in
\cite{leibbrandt} in the case of the light-cone gauge, showing that it
leads to an effect similar to the one observed in covariant gauges
\cite{shift,piguet,cmartin}. As in \cite{lcone}, we will assume that, at
most, higher loops account for a shift in the constant $k$.

The action (\ref{qaction}) have two important properties: it does not have
derivatives in the directions transverse to $x^0$, and it is quadratic in
the fields. This last property implies that vacuum expectation values can
be expressed in terms of the two-point correlation function. This
constitutes an important simplification of the analysis of the
perturbative series expansion associated to the vacuum expectation value
of the operators (\ref{operador}).

As pointed out in \cite{king,lcone}, in studying Chern-Simons gauge theory
in the light-cone it is convenient to Wick rotate the theory into
Euclidean space
$\RR\times \CC$. If one denotes a point in Euclidean space by 
$(t,z)$, where $z=x^1+i x^2$, after introducing $A_z = A_1 + i A_2$ and
$A_{\bar z} = A_1 - i A_2$ one finds \cite{king,lcone}:
\bear
\langle A_{\bar z}^a(x) A_m^b(x') \rangle &=& 0, \nonumber\\  
\langle A^a_m(x) A^b_n(x') \rangle &=& {4\pi\over k}\delta^{ab}
\epsilon_{mn} { 1 \over 2\pi i}  { \delta ( t - t'  )
\over z - z' },
\label{prop}
\eear  with $m,n = \{ 0,z \}$, and $\epsilon_{mn}$ is antisymmetric with 
$\epsilon_{0z}=1$. The Feynman rules of the theory are given by the form
of this propagator and the vertex associated to the Wilson lines:
\beq T^a \int dx^\mu,
\label{vertice}
\eeq where $T^a$ is a group generator and the integral is along the path
corresponding to one of the Wilson lines entering (\ref{operador}).

The perturbative series expansion emerging out of the application of the
Feynman rules corresponding to (\ref{prop}) and (\ref{vertice}) in the
computation of vacuum expectation values of Wilson loops was extensively
analyzed in \cite{lcone}. First, it was observed that due to the form of
the propagator (\ref{prop}) one must consider Morse knots. Similarly, in
the case under consideration, we will consider Morse singular knots where
no double points appear at the extrema. In addition, a framing was
introduced to deal with the problems generated by the singularities
inherent to the propagator  (\ref{prop}) when its end points coincide. For
the case of singular knots with no double points at their  extrema we can
proceed in the same way.

\begin{figure}
\centerline{\hskip.4in \epsffile{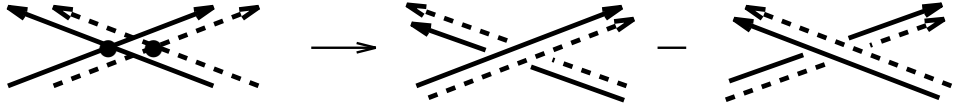}}
\caption{Resolution of a framed singular knot.}
\label{resol}
\end{figure}

We define framed singular knots considering the natural extension of
framed knots. In a framed singular knot with $n$ double points the knot
itself and its companion have $n$ self-intersections each. There is no
intersection between the knot and its companion. The resolution of a
double point of a framed singular knot is depicted in fig. \ref{resol}. 
For reasons that will become clear below, it is convenient to locate the
double points of the companion singular knot at the same heights as the
double points of the singular knot itself.

Framed singular knots induce a regularization which is achieved by
replacing a propagator (\ref{prop}) joining two lines by a propagator
which has one of its ends  attached to the singular knot itself and
the other to its companion. We must first prove that this regularization
renders the terms of the perturbative series expansion finite after the
zero-width limit is taken. But first we must recall some of the notation
introduced in \cite{lcone}, and reformulate it to suit the case of
singular knots.

Let us assume that the Morse singular knot $K^n$ possesses $2m$ extrema and
$n$ double points. In \cite{lcone},  a set of curves  joining the different
maxima and minima of $K^n$ where introduced. In the case of singular knots
we need to enlarge this set. If one of the curves joining two extrema goes
through a double point we will split it into two curves. We will consider
therefore $2(m+n)$ curves
$k^i$, $i=1,\dots,2(m+n)$, joining either two extrema,  an extremum and a
double point, or two double points. For each curve
$k^i$ there is a one-to-one correspondence among the points of $k^i$ and
the values that the variable $t$ takes. As in the non-singular case, $K^n$
can be regarded as a complex multivalued function of the variable $t$ with
$2(m+n)$ components. This observation implies that there exist an
alternative parametrization of $K^n$:
$z_i(t)$, $t\in I^i$, $i=1,\dots,2(m+n)$ where $z_i$ are the values that
the variable $z$ takes on $k^i$ and $I=[t_i^-,t_i^+]$ is the segment of
$\RR$ whose end-points are the values that the coordinate $t$ takes at the
two extrema or double points  joined by the curve $k^i$. With these data,
a propagator attached to the curves $k^i$ and $k^j$ can be expressed as
\cite{lcone}:
\beq  d x^\mu d x^\nu \langle A^a_\mu(x) A^b_\nu(x') \rangle 
\rightarrow {4\pi\over k} d t_i d t_j {1\over 2\pi i} \delta(t_i-t_j)
p_{ij} { \dot z_i(t_i) - \dot z_j(t_j) \over z_i(t_i) - z_j(t_j)},
\label{trading}
\eeq where:
\beq p_{ij} = 
\left\{  \begin{array}{lll} 
 1 & {\rm if} \,\, k_i \,\, {\rm and } \,\, k_j \,\, {\rm have} \,\, {\rm
the} \,\, {\rm same} \,\, {\rm orientation,}  &  \\  -1 &  {\rm if} \,\, 
k_i \,\, {\rm and } \,\, k_j \,\, {\rm have}  \,\, {\rm opposite} \,\,
{\rm orientations.} &  \end{array}
\right. 
\label{sign}
\eeq An analogous set of data, in which there are $2(m+n)$ curves
$k_\epsilon^i$, $i=1,\dots,2(m+n)$ and
$z$ is replaced by $z'$, is introduced for the companion knot.  After
performing one of the $t$-integrations, the regularization introduced by
the framing leads to the following formulae for the propagator
(\ref{trading}). For
$i=j$:
\beq {4\pi\over k}{1\over 2 \pi i}  {1\over 2} d s {\dot z_i(s) - \dot
z_i'(s) \over
                     z_i(s) - z_i'(s) },
\label{laurafer}
\eeq while for $i\neq j$:
\beq {4\pi\over k} {1\over 2 \pi i} {1\over 2} d s \Bigg(  {\dot z_i(s) -
\dot z_j'(s)
\over z_i(s) - z_j'(s) }+ {\dot z_i'(s) - \dot z_j(s) \over z_i'(s) -
z_j(s) } \Bigg) p_{ij}.
\label{martinfer}
\eeq

\begin{figure}
\centerline{\hskip.4in \epsffile{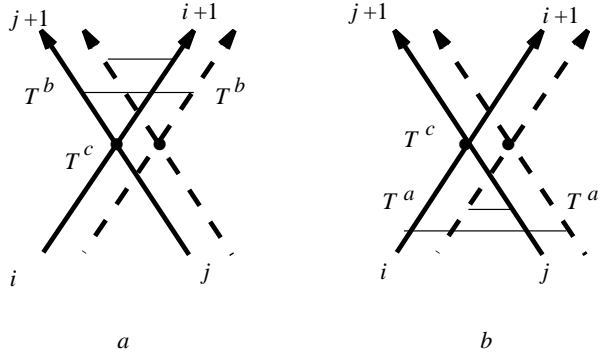}}
\caption{Propagator near a double point.}
\label{abobe}
\end{figure}

It was shown in \cite{lcone} that the regularization associated to framing
renders the perturbative series expansion finite in the zero-width limit
($\epsilon\rightarrow 0$) when the vacuum expectation value of a Wilson
loop is considered. The novelty when performing the same analysis for the
vacuum expectation values of the operators (\ref{operador}) is that now
there are  new divergences at the double points. However, the
regularization achieved by the introduction of a framing makes the sum
finite in the zero-width limit ($\epsilon\rightarrow 0$). The only
problematic contributions are those in which the two end-points of the
propagators are attached to the two lines involved in the double point.
There is always a contribution which approaches the double point from
below and another from above. These contributions have been pictured in
fig. \ref{abobe}  (the situation in which one of the directions is
inverted can be treated similarly). Only the part of the singular knot
near the double point has been depicted. First, notice that the group
factors corresponding to both situations are the same. Indeed, for diagram
$a$ one has a group factor,
\beq
\tr\big( T^c T^b X T^c T^b Y \big),
\label{phil}
\eeq while for diagram $b$:
\beq
\tr\big( T^c  X T^a T^c Y T^a \big),
\label{phildos}
\eeq where $X$ and $Y$ are some combinations of group generators which are
encountered as one travels along the singular knot. Certainly,
(\ref{phil}) and (\ref{phildos}) are the same group factor. The group
factors (\ref{phil}) and (\ref{phildos}) multiply an integrand which
contains a product of propagator contributions. The relevant factor near
the double point is given by the propagators shown in fig.
\ref{abobe}. If the double point takes place at height $t^*$, near the
double point we must deal with  integrals of the form:
\bear &&\int_{t^*}^{t_>} ds\,
\Bigg( {\dot z_{i+1}(s) - \dot z_{j+1}'(s) \over
       z_{i+1}(s) - z_{j+1}'(s)} +
      {\dot z_{i+1}'(s) - \dot z_{j+1}(s) \over
       z_{i+1}'(s) - z_{j+1}(s)} \Bigg)p_{i+1,j+1}, \nonumber\\  
&&\int^{t^*}_{t_<} ds\,
\Bigg( {\dot z_{i}(s) - \dot z_{j}'(s) \over
       z_{i}(s) - z_{j}'(s)} +
      {\dot z_{i}'(s) - \dot z_{j}(s) \over
       z_{i}'(s) - z_{j}(s)} \Bigg)p_{i,j}, 
\label{reino}
\eear where $t_>$ and $t_<$ are heights closed to $t^*$ such that $t_> >
t^*$ and $t_< < t^*$. Of course, the integrals in (\ref{reino}) are the
only part of the whole integral in which the contribution from the
diagrams in fig. \ref{abobe} differ. In the situation depicted in fig.
\ref{abobe}, both
$p_{i+1,j+1}=1$ and $p_{i,j}=1$ have value 1. If one of the directions
were inverted one would find $p_{i+1,j+1}=-1$ and $p_{i,j}=-1$. The
important fact is that in all cases $p_{i+1,j+1}=p_{i,j}$. The problematic
contributions from these integrals come from the region $s\rightarrow
t^*$. Carrying out the integration one finds:
\bear && \log\Bigg( { (z_{i+1}(t_>)-z_{j+1}'(t_>))
(z_{i+1}'(t_>)-z_{j+1}(t_>))
\over
 (z_{i+1}(t^*)-z_{j+1}'(t^*)) (z_{i+1}'(t^*)-z_{j+1}(t^*)) } \Bigg)
p_{i+1,j+1},
\nonumber\\   && \log\Bigg( { (z_{i}(t^*)-z_{j}'(t^*))
(z_{i}'(t^*)-z_{j}(t^*))
\over
 (z_{i}(t_<)-z_{j}'(t_<)) (z_{i}'(t_<)-z_{j}(t_<)) } \Bigg) p_{i,j}. 
\label{reinounido}
\eear The  denominator of the argument of the first logarithm as well as
the numerator of the argument of the second vanish in the zero-width
limit,
$\epsilon \rightarrow 0$, originating a logarithmic divergence. However,
for a finite value of $\epsilon$ these contributions cancel with each
other and one finds a finite answer for the sum in the zero-width limit.
Notice that this cancellation occurs if the height of the double point of
the companion singular knot is the same as the height of the singular knot
itself. Situations with a ladder of propagators approaching the double
point can be treated similarly. An equivalent picture that emerges from
this description is that at the double point one must use a principal value
prescription to avoid the singularity. We will describe this in more
detail below when dealing with an explicit example.

\begin{figure}
\centerline{\hskip.4in \epsffile{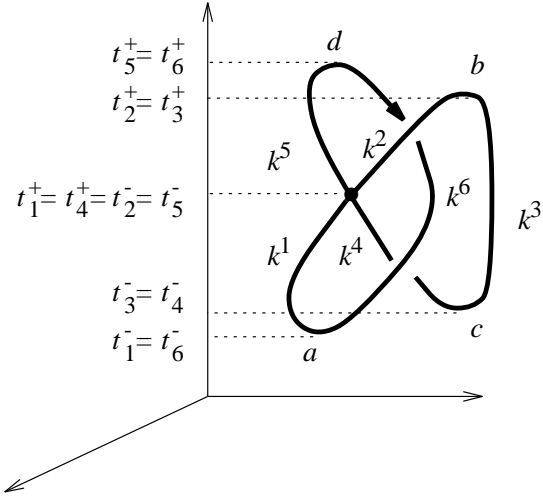}}
\caption{Example of a singular knot with a double point.}
\label{trsing}
\end{figure}

We are now in the position to write the general form of the terms of the
perturbative series expansion.  As in the case of non-singular knots, the
delta function in the height present in the propagator (\ref{prop})
implies that the only non-vanishing propagators are those in which their
two end-points have the same height. This observation allows to rearrange
the contributions to the perturbative series expansion in the following
way. Consider all possible pairings $\{z_i(s),z_j'(s)\}$ of curves $k^i$
and $k^j_\epsilon$, $i,j=1,\dots,2(m+n)$, where $2m$ is the number of
extrema  and $n$ is the number of double points. A contribution to the
vacuum expectation value of an operator of the form (\ref{operador}), at
order
$q$ in perturbation theory, involves a path-ordered integral in the
heights $s_1 < \dots < s_r < \dots < s_q$, of a product of $q$ propagators
(\ref{laurafer}) and (\ref{martinfer}) which contain,
\beq
\prod_{r=1}^q {d z_{i_r}(s_r) - d z_{j_r}'(s_r)\over z_{i_r}(s_r) -
z_{j_r}'(s_r)}.
\label{elproducto}
\eeq Products of this type are characterized by $q$ pairs of numbers
$(i_r,j_r)$ with $r=1,\dots,q$. We will call the ordered set of these $q$
pairs an ordered pairing and we will denote it by $P_q$. One must take
into account all the ordered pairings at each order in perturbation
theory. This means that one must sum over all the $P_q$.  The group factor
associated to each ordered pairing $P_q$ is easily obtained by first 
placing group-generators at the two ends of the propagators involved in
the ordered pairing, and then traveling along the kont building the
resulting trace of generators. Of course, when doing this in the case of
singular knots one encounters generators at the double points. We will
denote the resulting group factor by $R(P_q)$. As in the case of
non-singular knots we take care of all the signs originated from the
$p_{ij}$ in the propagator (\ref{martinfer}) introducing
\beq s(P_q) = \prod_{r=1}^q  p_{i_r j_r}.
\label{elsigno}
\eeq

\begin{figure}
\centerline{\hskip.4in \epsffile{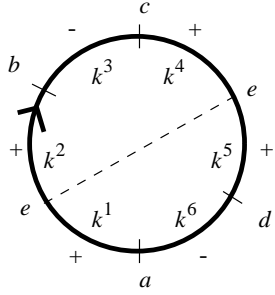}}
\caption{Representation of a singular knot with a double point on a
circle.}
\label{balon}
\end{figure}

The full expression for the contribution to the perturbative series
expansion at order $q$ takes the form:
\beq (2x)^{n+q}\Big({1\over 2\pi i}\Big)^q {1\over 2^q}\sum_{P_q}
\int_{t_{P_q}^-<t_1<\dots <t_r< \dots<t_m<  t_{P_q}^+} {\hskip-0.9cm} s(P_q)
\prod_{r=1}^q {d z_{i_r}(t_r) - d z_{j_r}'(t_r)\over z_{i_r}(t_r) -
z_{j_r}'(t_r)} R(P_q),
\label{eltung}
\eeq where $t_{P_q}^+$ and $t_{P_q}^-$ are highest and lowest heights
which can be reached by the last and first propagators of a given ordered
pairing $P_q$. This expression has a close resemblance to the one
presented in \cite{lcone} for the case of non-singular knots. There are,
however, three important differences. First, the contributions in
(\ref{eltung}) begin at order $n$. Second, in the path integrals in
(\ref{eltung}) the double points are regarded as sources of two extrema, a
maximum and a minimum. Third, the group factors $R(P_q)$ contain the group
generators placed at the double points.

\begin{figure}
\centerline{\hskip.4in \epsffile{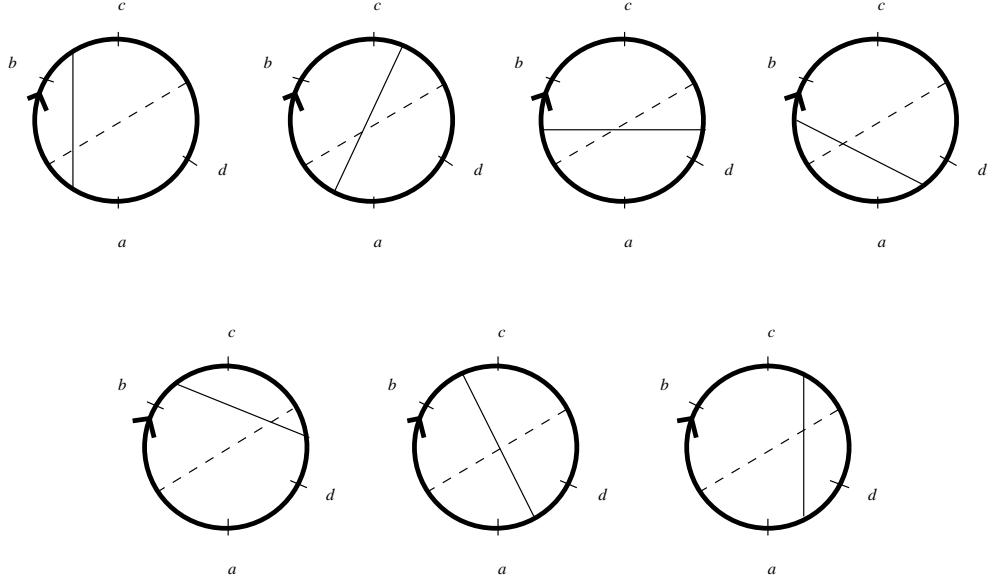}}
\caption{Contributions from the analysis on the associated circle.}
\label{balones}
\end{figure}

In order to illustrate the content of the integral (\ref{eltung}) we will
write down its explicit form for the singular knot depicted in fig.
\ref{trsing} at order $q=1$. Let us denote that singular knot by $K^1$ and
let us write down all the contributions to one of the group factors present
for $q=1$. In order to analyze the contributions we will proceed as in
\cite{lcone}. As shown in fig. \ref{trsing}, we must consider six curves,
$k^1,\dots,k^6$, whose end-points are extrema, $a$, $b$, $c$ and $d$, and
double points, $e$. The heights of these points are:
\bear
 a &\rightarrow & t_1^- = t_6^-, \nonumber \\
 b &\rightarrow & t_2^+ = t_3^+, \nonumber \\
 c &\rightarrow & t_3^- = t_5^-,  \\
 d &\rightarrow & t_5^+ = t_6^+, \nonumber \\
 e &\rightarrow & t_1^+ = t_4^+=t_2^-=t_5^-. \nonumber 
\label{abcde}
\eear  They are depicted in fig. \ref{trsing}. To obtain all contributions
we divide the circle which represents the singular knot in six parts taking
into account that the double point can be regarded as a maximum where the
curves
$k^1$ and
$k^4$ join, and a minimum where, similarly, the curves $k^2$ and $k^5$ get
attached. The resulting diagram has been depicted in fig. \ref{balon}.
Notice that a dashed line has been drawn to distinguish the points which
correspond to the double point. Also notice that a sign has been assigned
to each section of the circle. A plus indicates that the ascending
direction coincides with the knot direction. A minus indicates the
opposite.

\begin{figure}
\centerline{\hskip.4in \epsffile{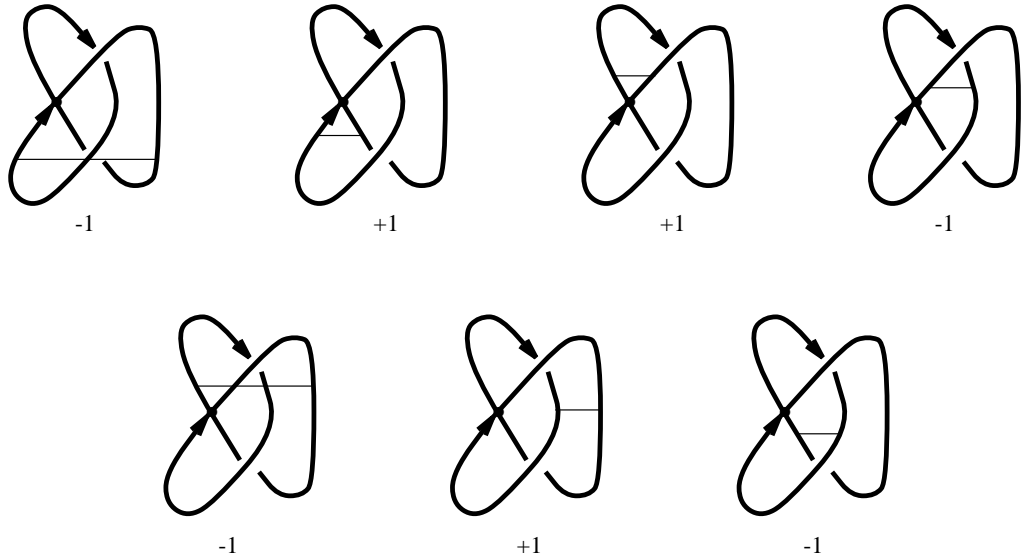}}
\caption{Contributions corresponding to a selected group factor.}
\label{trebs}
\end{figure}

The contributions at order $q=1$ are obtained from the diagrams resulting
after joining the curves with a propagator. The group factor for a given
choice is obtained  computing the trace which emerges after traveling
through the circle. To be specific, let us write the contributions
corresponding to the group factor $\tr(T^a T^b T^a T^b)$. The possible
contributions are encoded in the diagrams presented in fig. \ref{balones}.
In writing out all these contributions one must take into account that due
to the form of the propagator (\ref{prop}), a line can not be attached to
two sections of the circle  which are adjacent to a dashed line. The
contributions are easily depicted in the singular knot itself. For each
contribution one must compute a sign which is due to the product of the
$p_{ij}$ in (\ref{prop}). This sign is easily computed using the signs
introduced for the sections of the circle in fig. \ref{balon}. The
resulting signs, as well as the singular knot diagrams are pictured in
fig. \ref{trebs}. We will assume that the knot is located almost entirely
on a plane parallel to the time-axis. Only in a small neighborhood of the
crossings the singular knot gets slightly off the plane. 
The integral corresponding to each contribution is easily
obtained with the help of the general formula (\ref{eltung}). For the
first diagram one finds:
\beq -(2x)^2{1\over 2\pi i}{1\over 2}\int_{t_3^-}^{t_1^+} ds\,
\Bigg( {\dot z_1(s) - \dot z_3'(s) \over z_1(s)- z_3'(s) } +
       {\dot z_1'(s) - \dot z_3(s) \over z_1'(s)- z_3(s) }  \Bigg).
\label{primera}
\eeq This contribution is finite and framing independent so one can take
the zero-width limit before performing the integral. One finds:
\beq -(2x)^2{1\over 2\pi i} \log\Bigg( {z_1(t_1^+)-z_3(t_1^+) \over 
    z_1(t_3^-)-z_3(t_3^-)} \Bigg).
\label{priprima}
\eeq The contributions from the fourth to seventh diagrams  shown in fig.
\ref{trebs} can be obtained in the same way leading, respectively, to:
\bear && -(2x)^2{1\over 2\pi i} \log\Bigg( {z_2(t_2^+)-z_6(t_2^+) \over 
    z_2(t_2^-)-z_6(t_2^-)} \Bigg) -(2x)^2  {1\over 2}, \nonumber\\ &&
-(2x)^2{1\over 2\pi i} \log\Bigg( {z_5(t_3^+)-z_3(t_3^+) \over 
    z_5(t_5^-)-z_3(t_5^-)} \Bigg), \nonumber\\ && +(2x)^2{1\over 2\pi i}
\log\Bigg( {z_3(t_3^+)-z_6(t_3^+) \over 
    z_3(t_3^-)-z_6(t_3^-)} \Bigg), \nonumber\\ && -(2x)^2{1\over 2\pi i}
\log\Bigg( {z_4(t_4^+)-z_6(t_4^+) \over 
    z_4(t_4^-)-z_6(t_4^-)} \Bigg) -(2x)^2  {1\over 2}. \nonumber\\
\label{elresto}
\eear Notice that in the first and the fourth expressions of
(\ref{elresto}) we have taken into account the half turn occurring in the
fourth and seventh diagrams. In the remaining two contributions one must
be careful with the regularization  since each one is divergent in the
zero-width limit. From the second diagram we have,
\bear &&{\hskip-1cm} (2x)^2{1\over 2\pi i}{1\over 2}\int_{t_4^-}^{t_4^+}
ds\,
\Bigg( {\dot z_1(s) - \dot z_4'(s) \over z_1(s)- z_4'(s) } +
       {\dot z_1'(s) - \dot z_4(s) \over z_1'(s)- z_4(s) }  \Bigg)
 \nonumber\\ &&{\hskip-1cm} = (2x)^2{1\over 2\pi i} \log\Bigg(
{(z_1(t_4^+)-z_4'(t_4^+))  (z_1'(t_4^+)-z_4(t_4^+)) \over 
    (z_1(t_4^-)-z_4'(t_4^-)) (z_1'(t_4^-)-z_4(t_4^-))}\Bigg), 
\label{divuno}
\eear Similarly, the contribution from the third diagram is:
\beq (2x)^2{1\over 2\pi i} \log\Bigg( {(z_2(t_2^+)-z_5'(t_2^+)) 
(z_2'(t_2^+)-z_5(t_2^+)) \over 
    (z_2(t_2^-)-z_5'(t_2^-)) (z_2'(t_2^-)-z_5(t_2^-))}\Bigg).
\label{divdos}
\eeq Adding up (\ref{divuno}) and (\ref{divdos}) one gets a quantity which
is finite in the zero-width limit:
\beq (2x)^2{1\over 2\pi i} \log\Bigg( {z_2(t_2^+)-z_5(t_2^+)\over 
    z_1(t_4^-)-z_4(t_4^-) }\Bigg).
\label{divsum}
\eeq Summing all the contributions one finds a pairwise cancellation
obtaining finally, after putting back the group factor,
\beq -(2x)^2 \tr(T^a T^bT^aT^b) = (2x)^2 ({1\over 2} \sum_i \epsilon_i )
\tr(T^a T^bT^aT^b),
\label{cater}
\eeq where $\epsilon_i$, $i=1,2$, are the signs at each of the crossings
of one of the sections of the singular knot with the other section. This
result is a particular case of the general one which will be presented in
the next section for Vassiliev invariants of order $n$ corresponding to
singular knots with
$n-1$ double points.

The cancellation of the singularities at the double point suggests the
following equivalent picture  in which the framing is not taken into
account to avoid the divergences.  The idea is to avoid those divergences
introducing a principal value prescription. Let us consider a double point
as the one in fig. \ref{trsing} but forgetting about the companion knot
and let us introduce a small positive quantity $\eta$. The sum of the two
divergent integrals are regularized in the following form:
\beq
\int^{t^*-\eta}_{t_<}  d s {\dot z_i(s) - \dot z_j(s)\over z_i(s) -
z_j(s)}p_{ij} +
\int^{t_>}_{t^*+\eta} d s {\dot z_{i+1}(s) - \dot z_{j+1}(s)\over
z_{i+1}(s) - z_{j+1}(s)}p_{i+1,j+j}.
\label{finitodos}
\eeq This sum is finite in the limit $\eta\rightarrow 0$ and leads to the
same result as the one obtained regularizing using the framing. This
picture might result more useful in practical calculations.

\vfill
\eject

\section{Vassiliev invariants of order $n$ for singular knots with $n-1$
double points}
\setcounter{equation}{0}

Another direct application of the operators (\ref{operador}) is the
derivation of the Vassiliev invariants of order $n$ for singular knots with
$n-1$ double points. The result allows to fill very easily the next to top
row of an actuality table. 

Let us consider a singular knot $K^{n-1}$ with $n-1$ double points.  With
the help of the operators (\ref{operador}) one has that  the Vassiliev
invariants of order $n$ are given by  the following vacuum expectation
value:
\bear && {\hskip-1cm} v_n (K^{n-1}) = (2x)^{n-1} \Bigg\langle\tr
\Big[T^{\phi(w_1)} U(w_1,w_2) T^{\phi(w_2)} U(w_2,w_3) T^{\phi(w_3)}\cdots
\nonumber\\ && {\hskip3.5cm} \cdots U(w_{2n-3},w_{2n-2})
T^{\phi(w_{2n-2})} U(w_{2n-2},w_{1}) \Big]
\Bigg\rangle\Bigg|_{x^1},
\nonumber\\
\label{nextop}
\eear where the subindex $x^1$ indicates that one must keep only the terms
linear in $x$. To compute (\ref{nextop}) we must perform the
perturbative analysis of the theory. We will carry out the analysis in the
light-cone gauge using the construction presented in the previous section.
The same result is easily obtained in other gauges.

The simplifying feature of the first order analysis is that there are only
contributions from the crossings. This has been shown explicitly in the
example considered in the previous section but it is general. Actually, it
also holds in other gauges like the Landau gauge. The simplest way to see
this in that gauge is by considering the flat knot limit introduced in
\cite{alemanes}. The first order contribution is therefore very  simple,
one just has to count how many crossing sings contribute to each group
factor. 

\begin{figure}
\centerline{\hskip.4in \epsffile{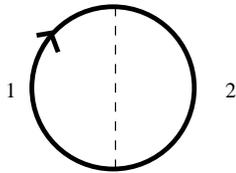}}
\caption{The only configuration for singular knots with one double point.}
\label{bola}
\end{figure}

 The analysis of the previous section indicates that we must consider the
singular knot as the union of a series of curves $k^i$ which join extrema
and double points. We observed a cancellation of the contributions at the
end points of these curves leaving only a part which counts the crossings
with signs. This will occur in general and since at the extrema there are
no insertions of generators it is most convenient to consider the singular
knot as a union of sections $l^j$, $j=1,\dots,2(n-1)$, for a knot with
$n-1$ double points. The reason for this is that contributions
corresponding to a propagator joining points of two different sections, or
points of a section with itself have a constant group factor as the
end-points move along. The contribution originated from a propagator
joining sections $l^j$ and $l^{j'}$ is
\bear&&{\hskip-1cm} (2x)^n {1\over 2}{\cal L}(j,j')\tr\Big[T^{\phi(w_1)}
T^{\phi(w_2)} \cdots T^{\phi(w_j)} T^a T^{\phi(w_{j+1})} \cdots \nonumber\\
&&{\hskip4cm}
\cdots T^{\phi(w_{j'})} T^a T^{\phi(w_{j'+1})} \cdots
T^{\phi(w_{2n-2})}\Big],
\label{belen}
\eear where ${\cal L}(j,j')$ is the linking number of the section $j$ with
section
$j'$,
\beq {\cal L}(j,j') = \sum_{i} \epsilon_i,
\label{enlace}
\eeq being the signs $\epsilon_i$ the ones in fig. \ref{signat}, and
the sum runs over all the crossings of section $j$ with section $j'$. In
the case that the end-points of the propagator are attached to the same
section of the singular knot one finds a similar contribution:
\beq (2x)^n {1\over 2}{\cal L}(j,j)\tr\Big[T^{\phi(w_1)} T^{\phi(w_2)}
\cdots T^{\phi(w_j)} T^a T^a T^{\phi(w_{j+1})} \cdots
T^{\phi(w_{2n-2})}\Big],
\label{belendos}
\eeq where ${\cal L}$ has the same form as in the other case,
\beq {\cal L}(j,j) = \sum_{i} \epsilon_i,
\label{enlacedos}
\eeq being now the signs $\epsilon_i$ the ones corresponding to the
crossings of the section $j$ with itself. Using (\ref{enlace}) and
(\ref{enlacedos}) one easily obtains the general form for $v_n (K^{n-1})$:
\bear&&{\hskip-1cm} v_n (K^{n-1})=(2x)^n {1\over 2}\sum_{j\leq j'} {\cal
L}(j,j')\tr\Big[T^{\phi(w_1)} T^{\phi(w_2)} \cdots T^{\phi(w_j)} T^a
T^{\phi(w_{j+1})} \cdots \nonumber\\ &&{\hskip6cm}
\cdots T^{\phi(w_{j'})} T^a T^{\phi(w_{j'+1})} \cdots
T^{\phi(w_{2n-2})}\Big],
\nonumber\\
\label{respu}
\eear where it is understood that for $j=j'$ this expression takes the
form (\ref{belendos}).

\begin{figure}
\centerline{\hskip.4in \epsffile{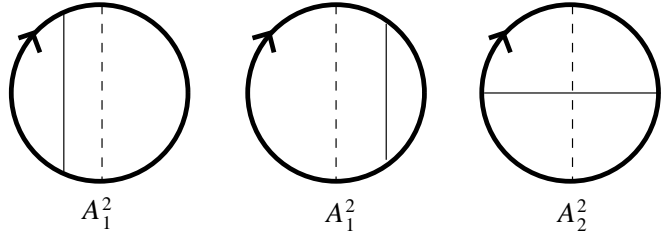}}
\caption{Diagrams corresponding to the contributions to $v_2(K^1)$.}
\label{bolatr}
\end{figure}

To illustrate the content of (\ref{respu}) we will present its  explicit
form for $n=2,3$. For the case $n=2$ we must consider singular knots with
one double point. There is only one possible configuration in this case.
It is pictured in fig. \ref{bola}. In order to obtain $v_2(K^1)$ as given
in (\ref{respu}) we must insert in the diagram shown in fig. \ref{bola} a
propagator in all possible ways. The three inequivalent possibilities are
shown in fig. \ref{bolatr}. The possible group factors at second order are:
\beq
A^2_1 = \tr (T^aT^aT^bT^b),
\,\,\,\,\,\,\,\,\,
A^2_2= \tr(T^aT^bT^aT^b).
\label{grupofa}
\eeq
Below each contribution shown in fig. \ref{bolatr} the corresponding group
factor is shown. The resulting form for  $v_2(K^1)$ is:
\beq 
v_2(K^1) = 2 x^2\Big[({\cal L}(1,1)+{\cal L}(2,2)) A_1^2 + {\cal
L}(1,2)A_2^2\Big].
\label{uvedos}
\eeq 

Before entering into the discussion of the Vassiliev invariants of order
three we will make a simple check of the result (\ref{uvedos}).  Using the
Lie-algebra relations of the generators, the expression (\ref{uvedos}) can
be written in terms of the group factors of a canonical basis
\cite{factor}:
\beq v_2(K^1) = 2 x^2 w(K^1) \tr\big( T^aT^aT^bT^b\big) + 2 x^2{\cal
L}(1,2) f_{abc}\tr\big(T^aT^bT^c\big), 
\label{uvedosotra}
\eeq where $w(K^1)$ is the writhe of the singular knot (the double point
does not contribute to $w(K^1)$). 

\begin{figure}
\centerline{\hskip.4in \epsffile{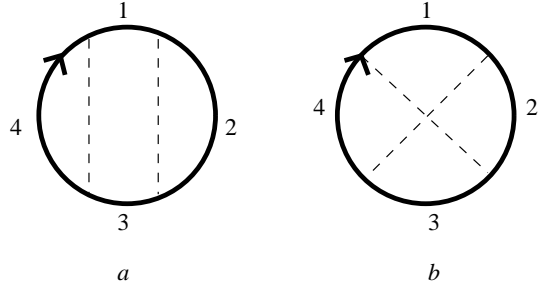}}
\caption{Inequivalent configurations of a singular knot with two double
points.}
\label{bolados}
\end{figure}

The result (\ref{uvedosotra}) can be verified very simply using the
explicit form for $v_2(K^0)$ given in \cite{alemanes}. In that paper it
was shown that:
\beq v_2(K^0) = {1\over 2} x^2 w(K^0)^2 \tr\big( T^aT^aT^bT^b\big) +
x^2\big(\rho_2(K^0) + \rho_1(K^0) \big) f_{abc} \tr\big(T^aT^bT^c\big),
\label{alema}
\eeq where,
\beq
\rho_2(K^0) = \sum_{(j>i)} \epsilon_i \epsilon_j,
\label{elrhodos}
\eeq being $\epsilon_i$ and $\epsilon_j$ the signatures of the crossings
$i$ and $j$ of $K^0$, respectively. By $(j>i)$ we mean pairs of crossings $i,j$ such that $s_i<s_j<t_i<t_j$. In (\ref{alema}), $w(K^0)$ is the
writhe of $K^0$ and $\rho_1(K^0)$ is a quantity which satisfies:
\beq
\rho_1(K^0_+) - \rho_1(K^0_-) = 0
\label{elrhouno}
\eeq where $K^0_+$ and $K^0_-$ differ by the exchange of an overcrossing
by an undercrossing. Using (\ref{elrhodos}) and (\ref{elrhouno}), as well
as the relations,
\bear w(K^0_+)^2 - w(K^0_-)^2 &=& 4 w(K^1), \nonumber\\
\rho_2(K^0_+) - \rho_2(K^0_-) &=& 2 {\cal L}(1,2),
\label{rela}
\eear one easily proves that, indeed, $v_2(K^0_+)-v_2(K^0_-)$ corresponds
to the expression obtained in (\ref{uvedosotra}).

\begin{figure}
\centerline{\hskip.4in \epsffile{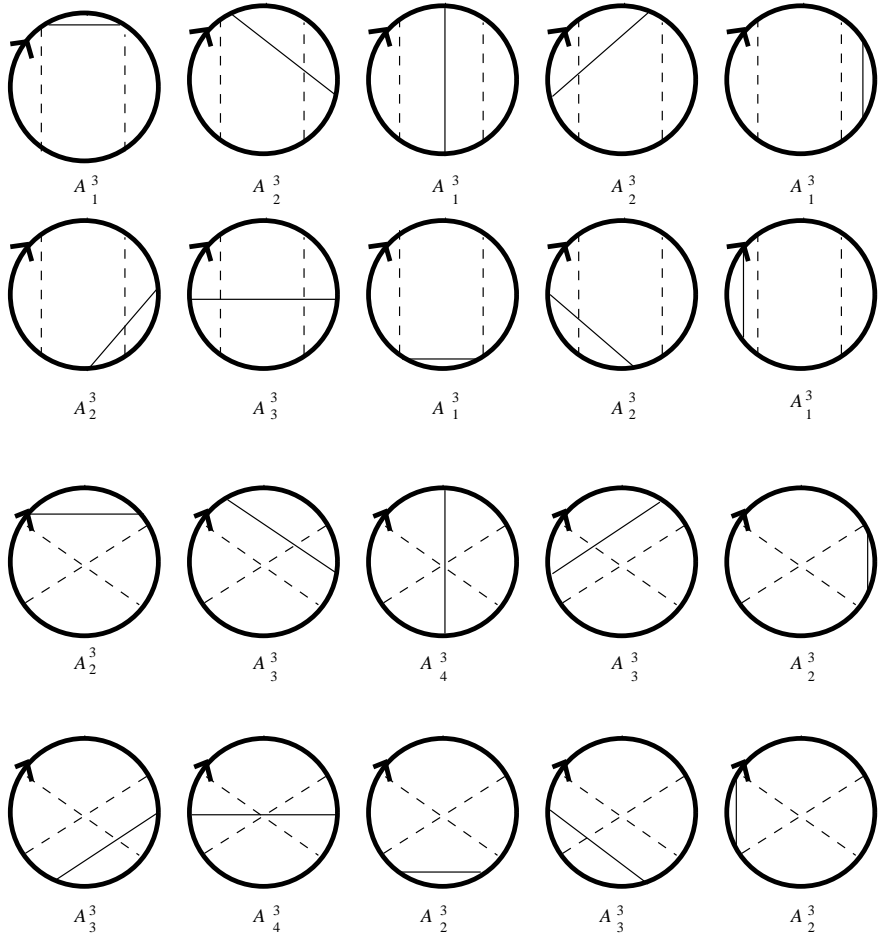}}
\caption{Diagrams corresponding to the contributions to $v_3(K^2_a)$ and
$v_3(K^2_b)$.}
\label{mubola}
\end{figure}

We will end this section presenting the explicit form of the third-order 
Vassiliev invariant for knots with two double points, $v_3(K^2)$. In this
case there are two inequivalent configurations of $K^2$. They are
presented in fig. \ref{bolados}. We will denote  by $K^2_a$ ($K^2_b$) a
choice of singular knot for the configuration $a$ ($b$). Our aim is to
compute $v_3(K^2_a)$ and $v_3(K^2_b)$ using (\ref{respu}). As before, we
begin labeling the sections of the singular knots as shown in fig.
\ref{bolados}, then we insert one propagator in all possible ways. To each
insertion corresponds a group factor. Denoting the possible group
factors at order three by:
\bear
A_1^3 &=& \tr (T^aT^aT^bT^bT^cT^c), \nonumber\\
A_2^3 &=& \tr (T^aT^bT^aT^bT^cT^c), \nonumber\\
A_3^3 &=& \tr (T^aT^bT^cT^aT^cT^b), \nonumber\\
A_4^3 &=& \tr (T^aT^bT^cT^aT^bT^c), 
\label{masgrupo}
\eear
one easily builds the table shown in fig. \ref{mubola}. With the help of
that table and (\ref{respu}) one finds:
\bear
&&{\hskip-1.4cm}
v_3(K^2_a) = 4 x^3 \Big[({\cal L}(1,1)+{\cal L}(2,2)+{\cal
L}(3,3) +{\cal L}(4,4)+{\cal L}(1,3)) A_1^3 \nonumber\\
&&{\hskip0.1cm}+ ({\cal L}(1,2)+{\cal
L}(1,4) +{\cal L}(2,3)+{\cal L}(3,4)) A_2^3 +{\cal L}(2,4) A_3^3 \Big],
\label{resa}
\eear
and,
\bear
&&{\hskip-1.3cm}
v_3(K^2_b) = 4 x^3 \Big[({\cal L}(1,1)+{\cal L}(2,2)+{\cal
L}(3,3) +{\cal L}(4,4)) A_2^3 \nonumber\\
&&
{\hskip-0.6cm}+ ({\cal L}(1,2)+{\cal L}(1,4) +{\cal L}(2,3)+{\cal L}(3,4))
A_3^3  +({\cal L}(1,3)+{\cal L}(2,4)) A_4^3 \Big]. \nonumber\\
\label{resb}
\eear

\begin{figure}
\centerline{\hskip.4in \epsffile{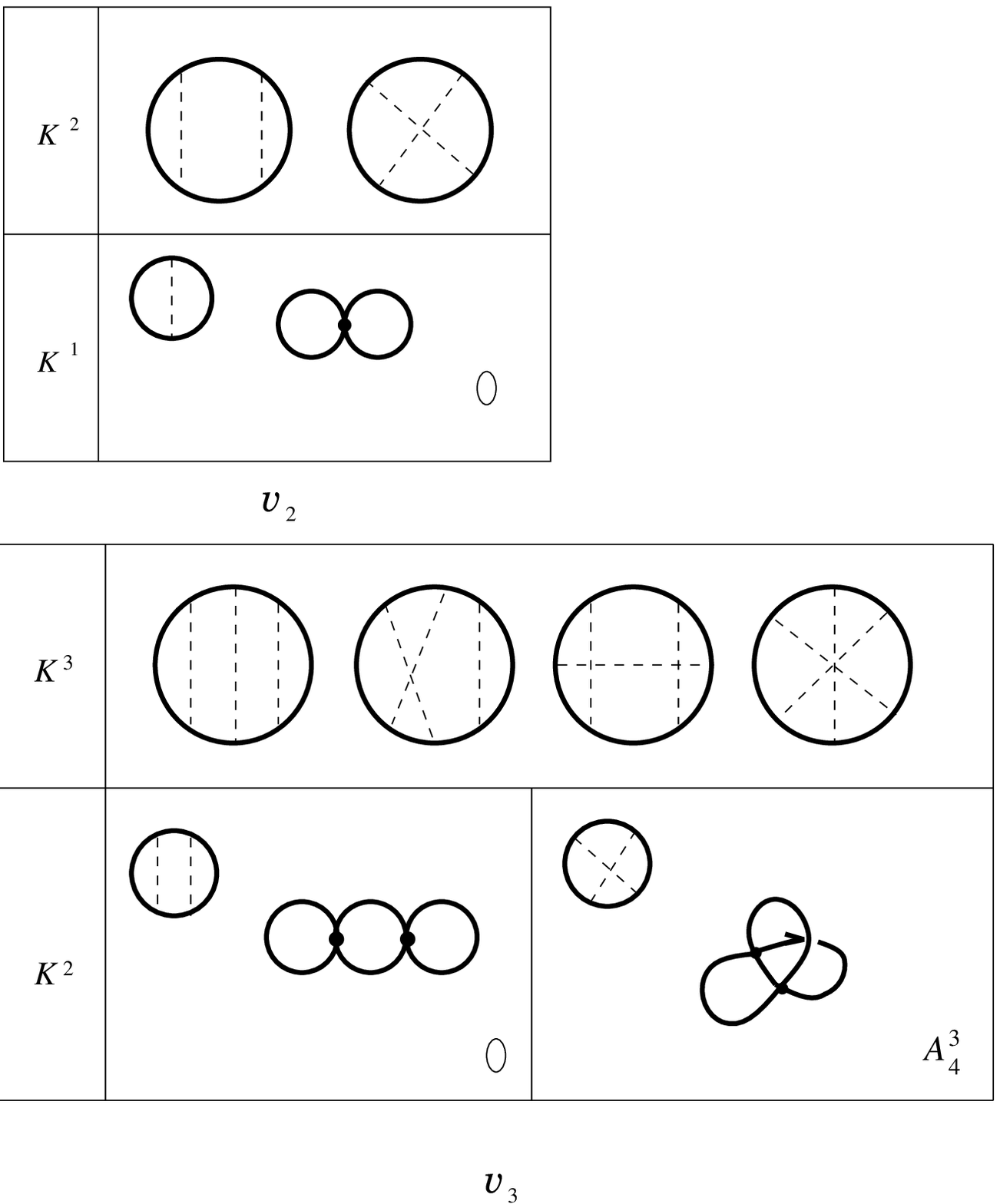}}
\caption{Elements of the actuality table for $v_2$ and $v_3$.}
\label{actua}
\end{figure}

The result (\ref{respu}) allows to complete in a very simple form the next
to the top row of an actuality table. In fig. \ref{actua} we have shown
some elements of the actuality tables corresponding to $n=2,3$.  Recall
that the top row is given by the form of the operators (\ref{operador}) at
order zero.  In the case of $n=2$ the table shown is the complete table.
Notice that in this case the two elements of the top row are independent.
For
$n=3$ only three of the four elements shown in top row are independent.
For the next to the top row we have chosen some representatives
corresponding to the two possible configurations and we have computed
(\ref{resa}) and (\ref{resb}) for each of them. The results are depicted
in the right lower corners. The next to the top row corresponding to
higher values of $n$ can be obtained similarly.

\vfill
\eject

\section{Conclusions}
\setcounter{equation}{0}

In this paper we have constructed the natural operators in the context of
Chern-Simons gauge theory to compute invariants associated to singular
knots. These operators have a very simple structure and consists of the
trace of an ordered series of Wilson lines, which are
associated to each of the sections of the singular knot, with adequate
insertions of group generators at the double points. They lead naturally
to the association of  configurations to singular knots in the standard
way in which it is done in knot theory. These operators allow to obtain
simple proofs of the Birman and Lin theorem for the expansion of a
polynomial invariant, and Bar-Natan's theorem for the integrability of
weight systems. Actually, our work constitutes a gauge independent (though
field-theory based) proof of this last theorem.

The analysis of Chern-Simons gauge theory in the light-cone gauge has
led to an expression for the vacuum expectation value of the operators
(\ref{operador}) which constitutes a generalization of the Knotsevich
integral to the case of framed singular knots. We have shown the
finiteness of this expression and we have computed it explicitly at lowest
order. This last result has provided the necessary information to obtain
the general form of the Vassiliev invariant of order $n$ for singular
knots with $n-1$ double points. We have shown how this expression allows
to construct very simply the next to the top row of an actuality table.

Our work opens a variety of investigations. One important issue consists
of the analysis of the operators (\ref{operador}) from a non-perturbative
point of view. Methods as the one used in  \cite{nbos,torus,martin,kaul}
should be applied to obtain information on the invariants associated to
singular knots. 

The study of the operators (\ref{operador}) from a perturbative
point of view should be pursued further. For example, it would be very
interesting to work out the structure of the perturbative series expansion
associated to the vacuum expectation values of the operators
(\ref{operador}) in covariant gauges. But perhaps the most important issue
that one should address within the perturbative approach is if the
operators (\ref{operador}) are enough powerful to provide relations among
Vassiliev invariants which could lead to a combinatorial formula for these
invariants, the simplest case being the one obtained for $v_n(K^{n-1})$.
These and other related issues will be addressed in future work.

\vskip2cm
\begin{center} {\bf Acknowledgements}
\end{center}

\vspace{4 mm}

 This work was supported in part by DGICYT under grant PB93-0344, and by
the EU Commission  under TMR grant FMAX-CT96-0012.

%
%
%
%

\end{document}